\documentclass[11pt,a4paper,useAMS,usenatbib]{emulateapj}
\usepackage{graphicx}
\usepackage{amsmath}
\usepackage{natbib}
\usepackage{epsfig}
\usepackage{pifont}

\def\lya{\ifmmode {\rm Ly}\alpha~ \else Ly$\alpha$~\fi}
\def\lyb{\ifmmode {\rm Ly}\beta~ \else Ly$\beta$~\fi}
\def\lyg{\ifmmode {\rm Ly}\gamma~ \else Ly$\gamma$~\fi}
\def\civ{\ifmmode {\rm C}\,{\sc iv}~ \else C\,{\sc iv}~\fi}
\def\civn{\ifmmode {\rm C}\,{\sc iv}~ \else C\,{\sc iv}\fi}
\def\cvn{\ifmmode {\rm C}\,{\sc v}~ \else C\,{\sc v}\fi}
\def\cvin{\ifmmode {\rm C}\,{\sc vi}~ \else C\,{\sc vi}\fi}
\def\nvn{\ifmmode {\rm N}\,{\sc v}~ \else N\,{\sc v}\fi}
\def\nvin{\ifmmode {\rm N}\,{\sc vi}~ \else N\,{\sc vi}\fi}
\def\nviin{\ifmmode {\rm N}\,{\sc vii}~ \else N\,{\sc vii}\fi}

\def\ovii{{{\rm O}\,\hbox{{\sc vii}}~}}
\def\oviii{{{\rm O}\,\hbox{{\sc viii}}~}}
\def\fexvii{{{\rm Fe}\,\hbox{{\sc xvii}}~}}
\def\neix{{{\rm Ne}\,\hbox{{\sc ix}}~}}
\def\nex{{{\rm Ne}\,\hbox{{\sc x}}~}}

\def\oviin{{{\rm O}\,\hbox{{\sc vii}}}}
\def\oviiin{{{\rm O}\,\hbox{{\sc viii}}}}
\def\neix{\ifmmode {\rm Ne}\,{\sc ix}~ \else Ne\,{\sc ix}~\fi}
\def\nex{\ifmmode {\rm Ne}\,{\sc x}~ \else Ne\,{\sc x}~\fi}
\def\hi{\ifmmode {\rm H}\,{\sc i}~ \else H\,{\sc i}~\fi}

\begin{document}

\title{Circumgalactic Medium on the Largest Scales: \\ Detecting X-ray Absorption Lines with Large-area Microcalorimeters\vspace{-1.3cm}}

\author{\'Akos Bogd\'an\altaffilmark{1},  Ildar Khabibullin\altaffilmark{2,3}, Orsolya E. Kov\'acs\altaffilmark{4}, Gerrit Schellenberger\altaffilmark{1}, John ZuHone\altaffilmark{1}, \\ Joseph N. Burchett\altaffilmark{5}, Klaus Dolag\altaffilmark{2,3}, Eugene Churazov\altaffilmark{3}, William R. Forman\altaffilmark{1}, Christine Jones\altaffilmark{1}, Caroline Kilbourne\altaffilmark{6}, Ralph P. Kraft\altaffilmark{1}, Erwin Lau\altaffilmark{1,7}, Maxim Markevitch\altaffilmark{6}, Dan McCammon\altaffilmark{8}, Daisuke Nagai\altaffilmark{9}, Dylan Nelson\altaffilmark{10}, Anna Ogorzalek\altaffilmark{6,11}, Benjamin D. Oppenheimer\altaffilmark{12}, Arnab Sarkar\altaffilmark{13}, Yuanyuan Su\altaffilmark{14}, Nhut Truong\altaffilmark{15}, Sylvain Veilleux\altaffilmark{9}, Stephan Vladutescu-Zopp\altaffilmark{2}, Irina Zhuravleva\altaffilmark{16}}  
\affil{\altaffilmark{1}Center for Astrophysics \ding{120} Harvard \& Smithsonian, 60 Garden Street, Cambridge, MA 02138, USA; abogdan@cfa.harvard.edu}
\affil{\altaffilmark{2}Universit\"ats-Sternwarte, Fakult\"at f\"ur Physik, Ludwig-Maximilians Universit\"at M\"unchen, Scheinerstr. 1, D-81679 M\"unchen, Germany}
\affil{\altaffilmark{3}Max Planck Institut f\"ur Astrophysik, Karl-Schwarzschild-Str.1, 85741 Garching bei M\"unchen, Germany}
\affil{\altaffilmark{4}Department of Theoretical Physics and Astrophysics, Faculty of Science, Masaryk University, Kotl\'a\v{r}sk\'a 2, Brno, 611 37, Czech Republic}
\affil{\altaffilmark{5} New Mexico State University, Department of Astronomy, Las Cruces, NM 88001, USA}
\affil{\altaffilmark{6} NASA Goddard Space Flight Center, X-ray laboratory, Greenbelt, MD 20771, USA}
\affil{\altaffilmark{7} Department of Physics, University of Miami, Coral Gables, FL 33124, USA}
\affil{\altaffilmark{8} Department of Physics, University of Wisconsin, Madison, WI 53706, USA}
\affil{\altaffilmark{9} Department of Physics, Yale University, New Haven, CT 06520, USA}
\affil{\altaffilmark{10} Universit\"at Heidelberg, Zentrum f\"ur Astronomie, Institut f\"ur theoretische Astrophysik, Albert-Ueberle-Str. 2, 69120 Heidelberg, Germany}
\affil{\altaffilmark{11} Department of Astronomy, University of Maryland, College Park, MD 20742, USA}
\affil{\altaffilmark{12} 5CASA, Department of Astrophysical and Planetary Sciences, University of Colorado, 389 UCB, Boulder, CO 80309, USA}
\affil{\altaffilmark{13} Kavli Institute for Astrophysics and Space Research, Massachusetts Institute of Technology, 77 Massachusetts Ave, Cambridge, MA
02139, USA, Cambridge, MA 02138, USA}
\affil{\altaffilmark{14} University of Kentucky, 505 Rose street, Lexington, KY 40506, USA}
\affil{\altaffilmark{15} Max-Planck-Institut f\"ur Astronomie, K\"onigstuhl 17, 69117 Heidelberg, Germany}
\affil{\altaffilmark{16} Department of Astronomy and Astrophysics, The University of Chicago, Chicago, IL 60637, USA}
\shorttitle{Detecting the CGM in Absorption}
\shortauthors{BOGD\'AN ET AL.}

\begin{abstract}
The circumgalactic medium (CGM) plays a crucial role in galaxy evolution as it fuels star formation, retains metals ejected from the galaxies, and hosts gas flows in and out of galaxies. For Milky Way-type and more massive galaxies, the bulk of the CGM is in hot phases best accessible at X-ray wavelengths. However, our understanding of the CGM remains largely unconstrained due to its tenuous nature. A promising way to probe the CGM is via X-ray absorption studies. Traditional absorption studies utilize bright background quasars, but this method probes the CGM in a pencil beam, and, due to the rarity of bright quasars, the galaxy population available for study is limited. Large-area, high spectral resolution X-ray microcalorimeters offer a new approach to exploring the CGM in emission and absorption. Here, we demonstrate that the cumulative X-ray emission from cosmic X-ray background sources can probe the CGM in absorption. We construct column density maps of major X-ray ions from the {\it Magneticum} simulation and build realistic mock images of nine galaxies to explore the detectability of X-ray absorption lines arising from the large-scale CGM. We conclude that the \ovii absorption line is detectable around individual massive galaxies at the $3\sigma-6\sigma$ confidence level. For Milky Way-type galaxies, the \ovii and \oviii absorption lines are detectable at the $\sim\,6\sigma$ and $\sim\,3\sigma$ levels even beyond the virial radius when co-adding data from multiple galaxies. This approach complements emission studies, does not require additional exposures, and will allow probing of the baryon budget and the CGM at the largest scales.
\end{abstract}


\section{Introduction}
\label{sec:intro}

The circumgalactic medium (CGM) is a critical component of galaxies and plays a key role in their evolution in concert with the supermassive black hole, the stellar component, and the dark matter halo. The CGM is defined as the multi-phase gas residing beyond the stellar component and within the virial radius of galaxies \citep[see][for a review]{tumlinson17}. The importance of this component is underscored by the fact that cooling of the CGM provides fuel for star formation in the galaxies, it is the repository of metals ejected from galaxies by feedback from supernovae (SNe) and active galactic nuclei (AGN), and it is the site where gas flows in and out of galaxies \citep[e.g.][]{tumlinson11,thom12,stocke13, peeples14,werk14,borthakur15,bogdan23}. Because the CGM retains signatures of all critical physical processes that shape the evolution of galaxies, studying and characterizing the large-scale CGM can resolve many major questions about galaxy evolution. 

Observations of the CGM demonstrate that this gas is multiphase and has a complex temperature and ionization structure. The cooler phases of the CGM are routinely explored in the ultraviolet wavelength using absorption studies. Observations of low-redshift galaxies using Hubble's Cosmic Origin Spectrograph demonstrated the feasibility of this approach and established the metal budget and characterized the physical properties of the cooler phases \citep[e.g.][]{werk14,peeples14,prochaska17}. However, the bulk of the CGM of Milky Way-type and more massive galaxies lies in the hot, X-ray emitting phases that can be best probed at X-ray wavelengths \citep{nelson18}. While studies using the Sunyaev-Zeldovich (SZ) signal of galaxies attempted to detect the hot gas around galaxies \citep{planck13,greco15,bregman22}, due to the weak signal from individual galaxies, these analyses must resort to stacking large samples of galaxies. These works revealed a signal for stacks of massive galaxies ($M_{\rm \star} \gtrsim (2-3) \times10^{11} \ \rm{M_{\odot}}$), but Milky Way-like galaxies remained undetected. Additionally, due to the relatively low spatial resolution of SZ measurements, these studies can only probe the ensemble of the emission from the interstellar- and circumgalactic medium and cannot probe the spatial structure of the gas.  

The hot X-ray emitting phases of the gas have been studied for decades with all major X-ray observatories. The first detections were obtained using the Einstein observatory \citep[e.g.][]{forman85}; observations of massive elliptical galaxies demonstrated the presence of hot gas extending beyond the stellar light. While detailed studies with \textit{Chandra} and \textit{XMM-Newton} characterized the gas around the most massive ellipticals, it is important to realize that most of these galaxies reside in a rich cluster or group environments. These larger dark matter halos have their own large-scale gas atmospheres that cannot be differentiated from the CGM associated with galaxies. X-ray observations were less successful in studying the CGM of isolated spiral galaxies. The CGM of massive ($M_{\rm \star} \gtrsim3\times10^{11} \ \rm{M_{\odot}} $) spiral galaxies were detected to a fraction ($\sim15\%$) of their virial radius around a handful of massive galaxies \citep{anderson11,bogdan13a,bogdan13b,anderson16,bogdan17,li17}. These observations provided limited measurements of the thermodynamic properties and spatial structure of the gas within about $60$~kpc, but the large-scale properties ($\gtrsim0.15R_{\rm 200}$) of the gas remained completely unconstrained.  While recent stacking studies of \textit{eROSITA} X-ray data hinted at the presence of extended emission around disk galaxies \citep{chadayammuri22,comparat22}, observational constraints on individual Milky Way-type galaxies are completely lacking \citep[see also][]{bogdan22}. 

The difficulty in detecting large-scale diffuse X-ray emission around individual galaxies stems from the low temperature and density of the hot gas and the comparatively high level of emission associated with the Milky Way foreground \citep{mccammon02}. Combining these effects with the relatively poor spectral resolution of traditional Charge Coupled Device (CCD) detectors, it becomes apparent that mapping low surface brightness features are not feasible with the present instrumentation. A potential way forward is to study absorption lines that are imprinted on the spectrum of a luminous bright background source, typically a bright quasar. This method has been used to probe the warm-hot intergalactic medium in the filamentary structure of the universe with varying success \citep[e.g.][]{nicastro05,kaastra06,nicastro18,johnson19,kovacs19}. Moreover, this approach could be applicable for studying the CGM around individual galaxies albeit no detection was achieved thus far \citep{yao10}. The disadvantages of this method are twofold. First, the number of galaxies that can be studied by such absorption studies is limited, and, for stacking studies, the observed result corresponds to the average of individual galaxies, whose properties and environment may be poorly constrained. Second, since absorption studies represent a pencil beam and the CGM is believed to exhibit significant spatial variation, the average properties of the CGM cannot be recovered. 

In this work, we introduce a new method that offers a viable and robust approach to studying the large-scale CGM of galaxies with a wide range of stellar masses. This method promotes a wide-field X-ray integral field unit (IFU) that simultaneously offers the imaging capabilities of CCDs and the high spectral resolution of grating instruments. Thanks to the large field of view and sufficiently deep observations, it will become possible to use the combined emission of cosmic X-ray (CXB) background sources and study the CGM out to the virial radius and beyond in absorption. The advantage of this method is that by using a substantial number of CXB sources, we will not only collect information about a (random) sightline through the CGM but will get a representative picture of the entire gaseous halo. Because the CXB sources are distributed approximately uniformly across the sky, not only selected galaxies with bright background quasars can be studied, but any nearby galaxy can be probed. This, in turn, will allow studying a representative sample of individual galaxies, while also boosting the signal-to-noise ratios by further stacking the data of individual galaxies.  

This paper is structured as follows. In Section 2 we describe the basic parameters of the nominal instrument design and the analyzed simulations used in this work. The analysis of the mock images and spectra is described in Section 3. We present the results on massive and Milky Way-type galaxies in Section 4. Our results are placed in context and are discussed in Section 5. Section 6 summarizes the results.

\begin{figure*}
  \begin{center}
      \includegraphics[width=0.9\textwidth]{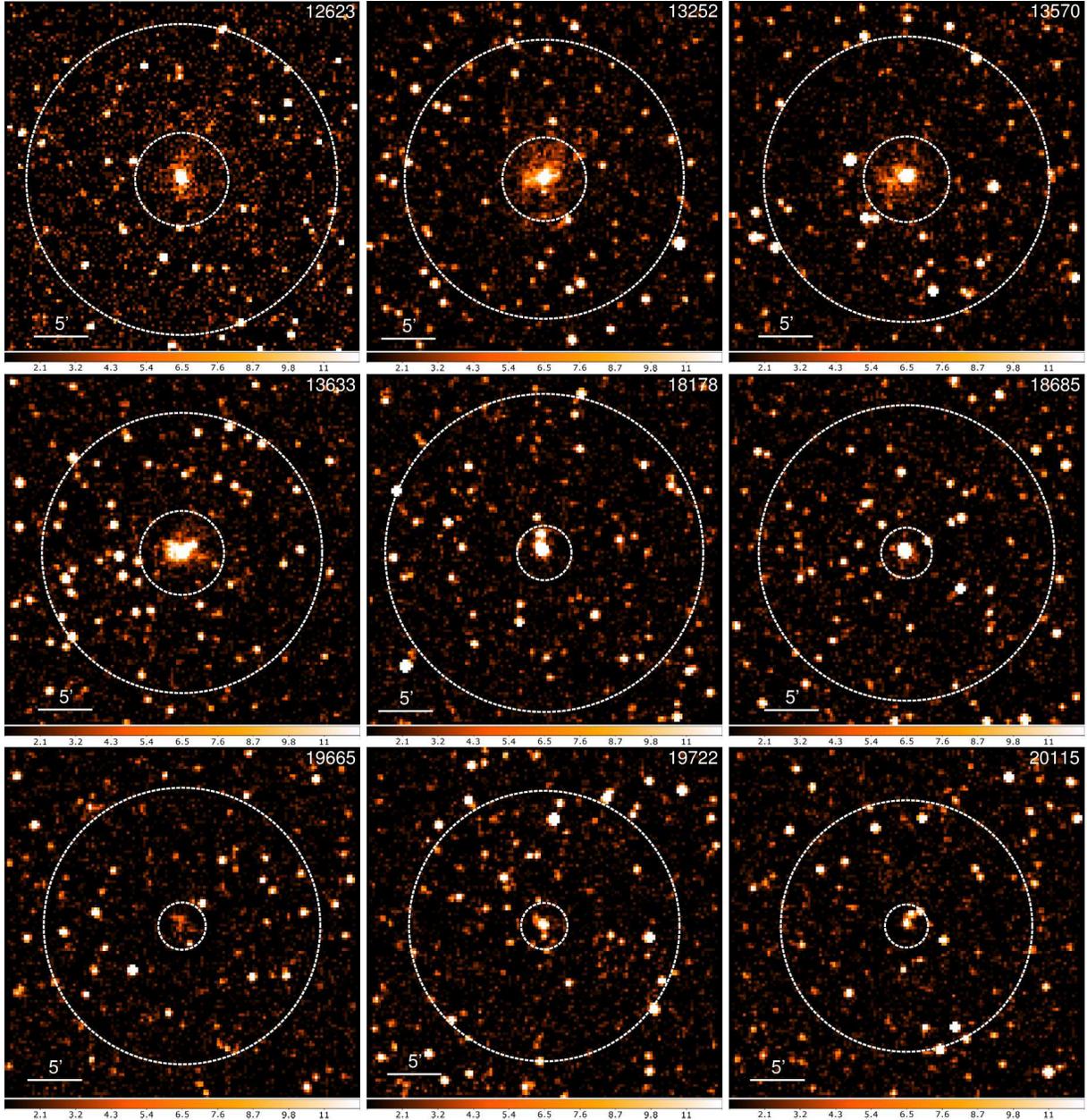}
      \caption{LEM mock X-ray images of the nine {\it Magneticum} galaxies studied in this paper. The galaxies are placed at $z=0.035$, at which redshift the absorption lines are shifted away from the Milky Way foreground lines. The images are the sum of two 2~eV wide images around the redshifted \ovii forbidden and resonance lines. The mock images include the main X-ray emitting components: (1) the emission from the CGM, (2) the Milky Way foreground emission including the Local Hot Bubble and the Hot Halo component, and (3) the population of CXB sources with a different realization for each galaxy. The images were scaled to highlight the CXB source population. The annuli show the $R=(0.3-1)R_{\rm 200}$ region for massive galaxies (first four panels) and the $R=(0.3-1.75)R_{\rm 200}$ for the Milky Way-type galaxies (last five panels).} 
     \label{fig:images}
  \end{center}
\end{figure*}

\begin{figure*}
  \begin{center}
      \includegraphics[width=0.95\textwidth]{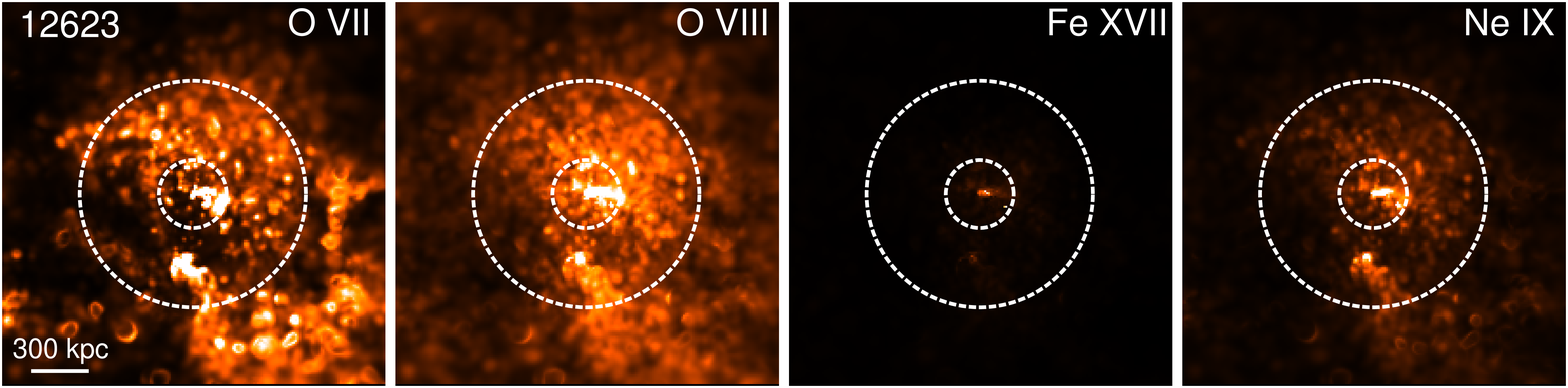}
      \includegraphics[width=0.95\textwidth]{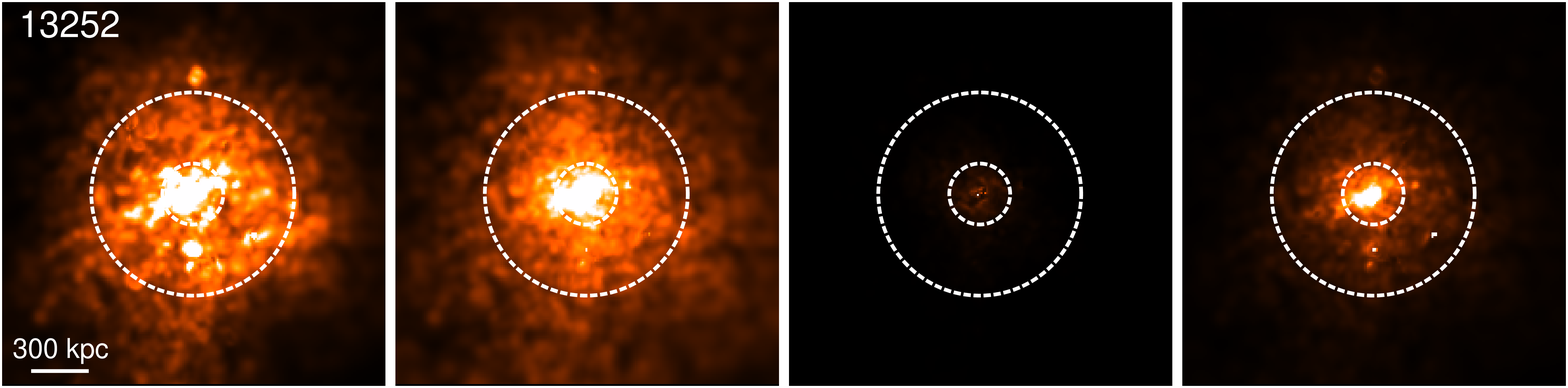}
      \includegraphics[width=0.95\textwidth]{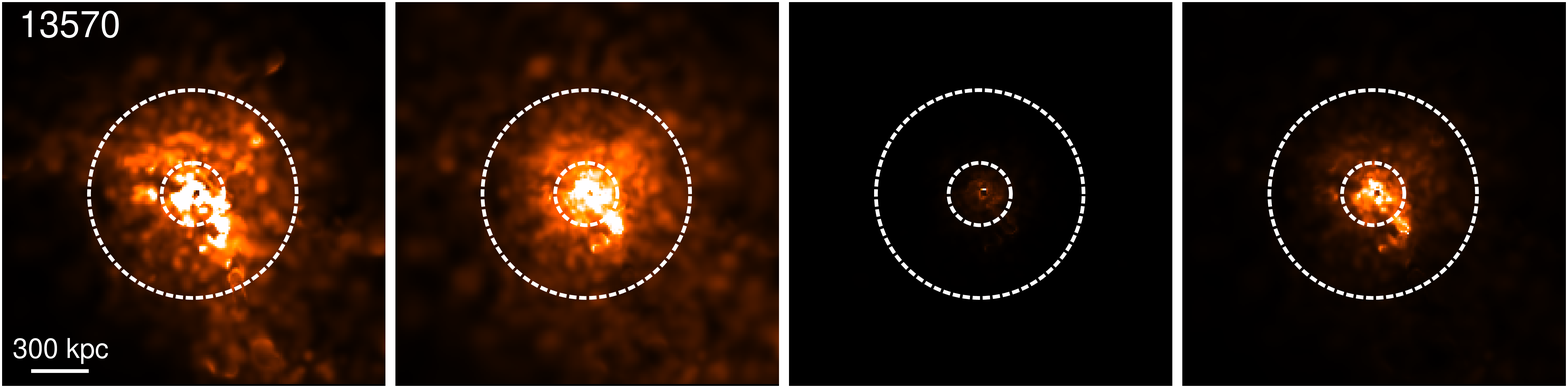}
      \includegraphics[width=0.95\textwidth]{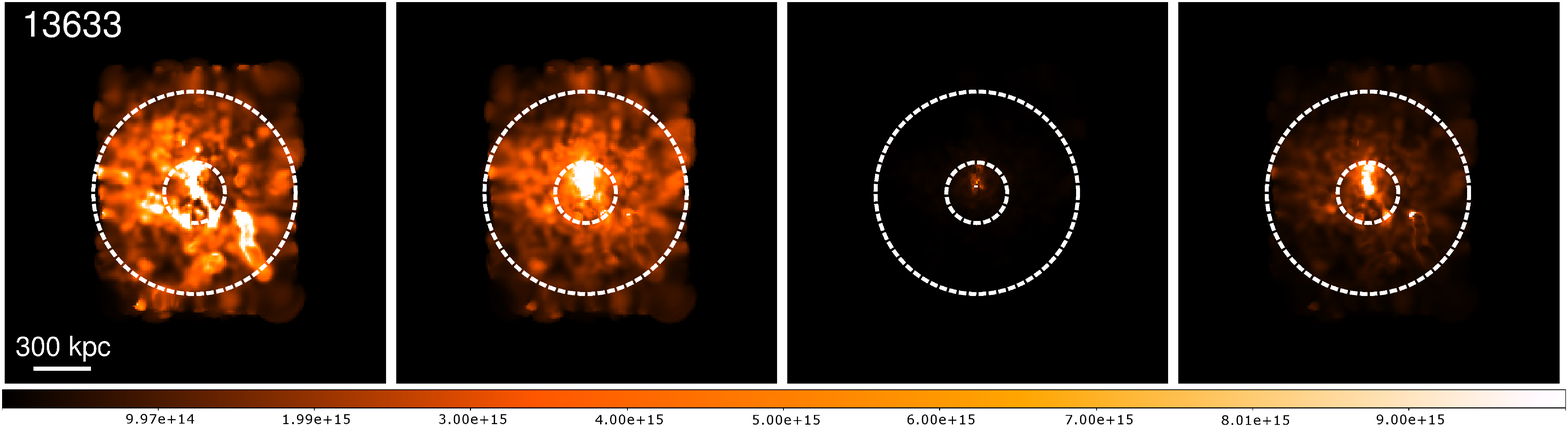}
      \caption{Column density maps of four massive galaxies with $M_{\rm 200} = (0.94-1.29) \times 10^{13} \ \rm{M_{\odot}} $ for the {\it Magneticum} simulation for the major ions. The annulus shows the $(0.3-1)R_{\rm 200}$ region of each galaxy (Table \ref{tab:galaxies}). The resolution of the maps is 10~kpc per pixel and each image has a side length of 2~Mpc. The scaling of the images is the same, shown with the color bar at the bottom panel. Note that while the column densities of \ovii and \oviii are comparable, the column density of \fexvii and \neix is nearly an order of magnitude lower. The lack of a clear signal in the \fexvii maps is due to their low column densities.}  
     \label{fig:nhmaps_massive}
  \end{center}
\end{figure*}

\begin{figure*}
  \begin{center}
      \includegraphics[width=0.95\textwidth]{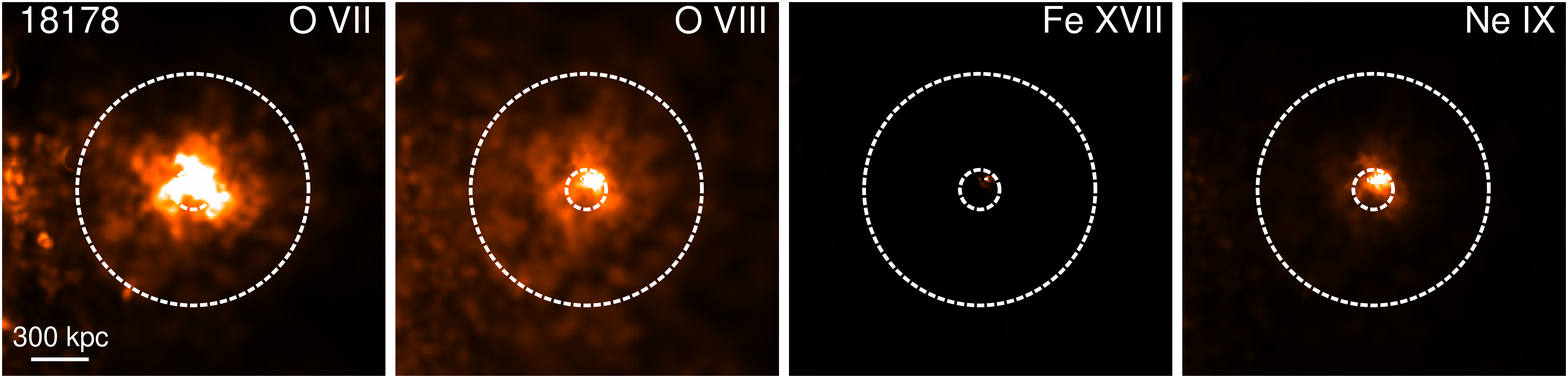}
      \includegraphics[width=0.95\textwidth]{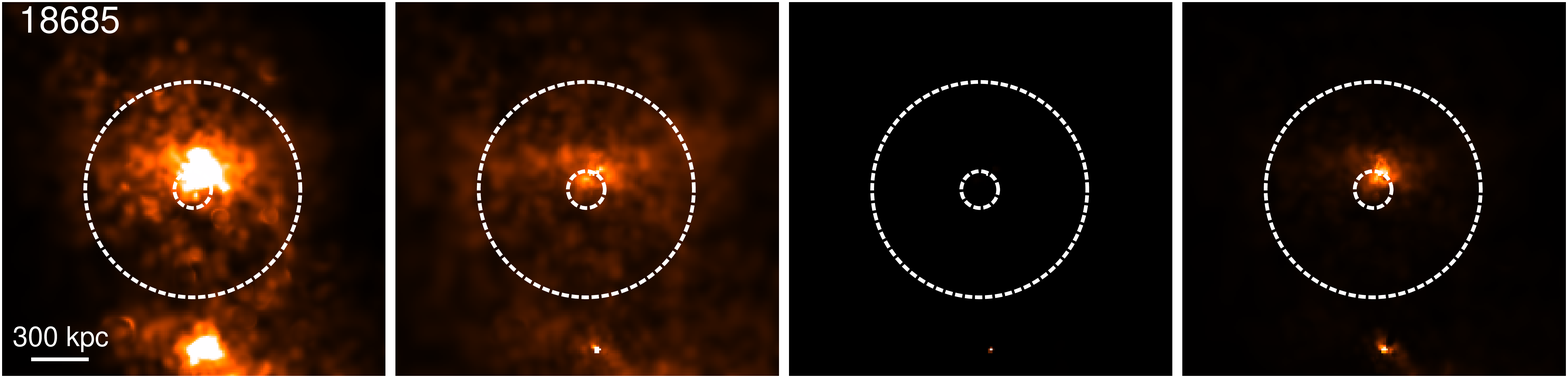}
      \includegraphics[width=0.95\textwidth]{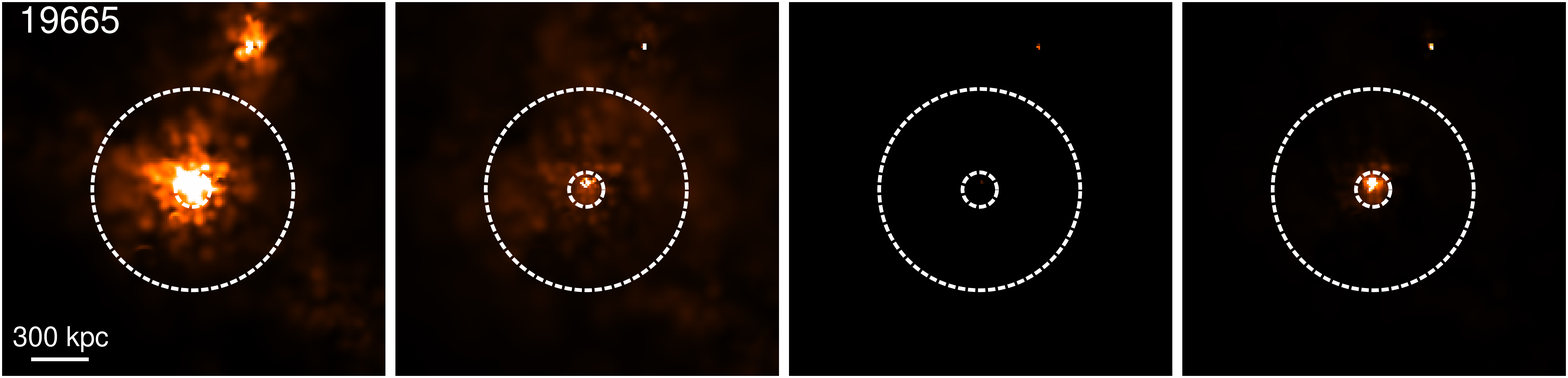}
      \includegraphics[width=0.95\textwidth]{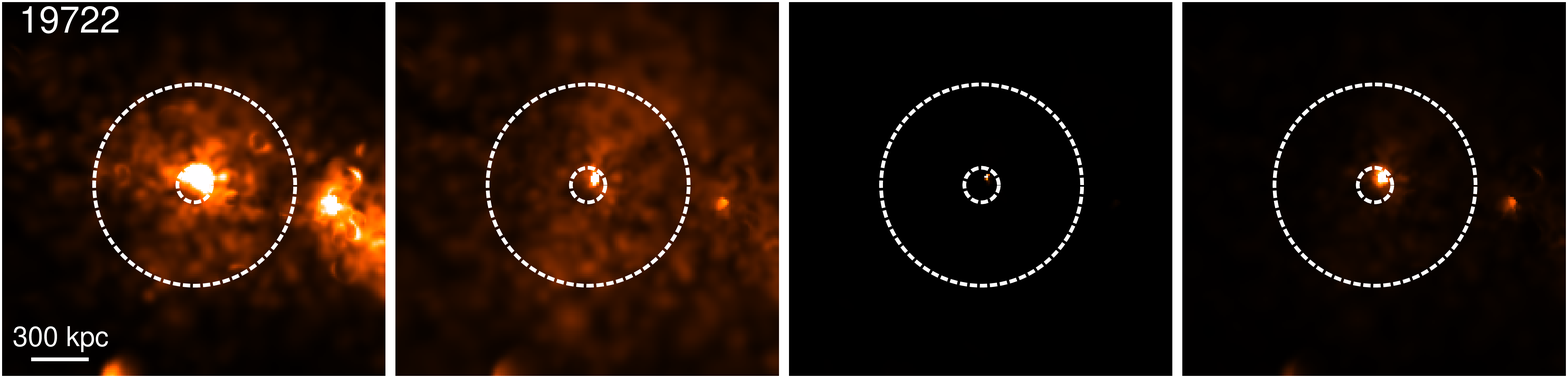}
      \includegraphics[width=0.95\textwidth]{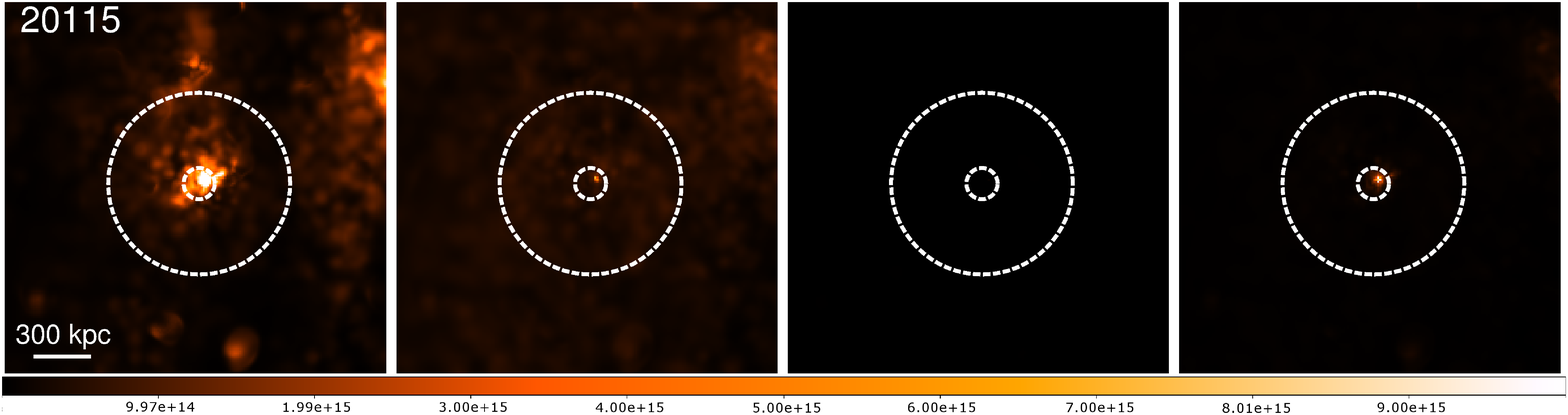}
      \caption{Same as Figure \ref{fig:nhmaps_massive}, but for Milky Way-type galaxies with $M_{\rm 200} = (1.27-2.57) \times 10^{12} \ \rm{M_{\odot}}$. The annulus shows the $(0.3-1.75)R_{\rm 200}$ region of each galaxy (Table \ref{tab:galaxies}). The column densities are notably lower than those for massive galaxies.} 
     \label{fig:nhmaps_lowmass}
  \end{center}
\end{figure*}

\section{Methods}
\label{sec:methods}

\subsection{The instrument parameters}
\label{sec:instrument}

In this study, we explore the feasibility of a new approach to detect X-ray absorption lines in the large-scale CGM of galaxies. To detect absorption lines in the cumulative spectrum of CXB sources, a high-resolution X-ray IFU is needed, such as the Athena X-IFU or the LEM microcalorimeter. While the results presented in this work can be achieved with \textit{Athena} X-IFU measurements (see Section \ref{sec:instruments}), here we opt to use the notional design parameters of the LEM mission concept \citep{kraft22}. The main reason for this choice is motivated by the larger field of view of LEM, which allows studying the CGM of nearby galaxies to virial radius in a single pointing. 

In this work, we assumed that the X-ray IFU has a $30\arcmin \times 30\arcmin $ field of view and a collecting area of $\approx1600 \ \rm{cm^2}$ at 0.5~keV energy. Because LEM is designed to be a soft X-ray instrument, it is sensitive in the $0.2-2$~keV band, which is ideal to detect the CGM of nearby galaxies. The collecting area as a function of energy (i.e.\ the effective area curve) is described in \citet{kraft22}. The product of the field of view and the effective area is referred to as the ``grasp'', which is $1.4\times10^6 \ \rm{cm^2 \ arcmin^2}$ at 0.5~keV for LEM, and surpasses the grasp of other planned or proposed instruments, such as XRISM ($5\times10^2 \ \rm{cm^2 \ arcmin^2}$) or Athena X-IFU ($1.2\times10^5\ \rm{cm^2 \ arcmin^2}$), by orders of magnitude\footnote{The grasp for Athena corresponds to the pre-reformulation requirements.}. As we discuss in Section \ref{sec:instruments} this analysis will be feasible with the Athena X-IFU instruments albeit the required exposure times will be about substantially longer. 

The angular resolution of the LEM concept is $10\arcsec$ (half-power diameter), which is critical to disentangle the CGM emission from bright background sources. For this work, the importance of the angular resolution is that we can use relatively small regions around the CXB sources, which regions will include $>90\%$ of the counts associated with the sources. This, in turn, allows us to minimize the contribution of other X-ray emitting sources, such as emissions from the Milky Way foreground or the galaxy CGM itself. 

The key enabling technology of the LEM mission concept and the \textit{Athena} X-IFU is the superb spectral resolution thanks to their state-of-the-art microcalorimeter array. The notional design of LEM has a 2~eV spectral resolution across the field of view with an $8\arcmin \times 8\arcmin$ subarray with 0.9~eV resolution in the center. This spectral resolution is better than that of XRISM (5~eV) and is comparable to that of \textit{Athena} X-IFU (2.5~eV). The high spectral resolution is essential to resolve narrow lines and line blends in the line-rich soft band. In this work, we do not consider the LEM subarray with its sub-eV resolution as the bulk of the emission from the large-scale CGM will fall beyond this region. For example, at $z=0.035$ the subarray corresponds to a region with $335\ \rm{kpc} \times 335\ \rm{kpc}$ region (or $167.5$~kpc in radius for a circular region), which is about $30\%$ of the virial radius of the massive Magneticum galaxies in our sample. Moreover, the CGM within this region can be well studied in emission, which is the subject of a parallel study (Schellenberger et al. in prep; Truong et al. in prep). Because we aim to detect the absorption signal beyond $0.3R_{\rm 200}$, omitting the 0.9~eV spectral resolution subarray is a reasonable and conservative choice.

\subsection{Mock X-ray images}
\label{sec:mocks}

In Figure \ref{fig:images} we present the narrow band mock images of the nine {\it Magneticum} galaxies that we study in this work. These images are the sum of two 2~eV wide images corresponding to the \ovii forbidden ($E=560.98$~eV) and resonance lines ($E=573.95$~eV) redshifted with z=0.035. The choice of the $z=0.035$ redshift was motivated by two main factors. First, at this redshift, the absorption signal from the CGM, most notably the \ovii resonance line, is shifted away and is cleanly separated from the Milky Way foreground lines. Second, the virial radius of galaxies at this redshift approximately fills the $30\arcmin\times30\arcmin$ field of view of the instrument. These X-ray images include all components that play a notable role in the X-ray appearance of the galaxies. These components are the large-scale gaseous CGM emission associated with the galaxies, the Milky Way foreground emission, and the emission from the CXB sources. The detailed spectral models of each of these components is presented in Section \ref{sec:data}.

To generate the mock X-ray data, we followed the method described in previous works \citep[e.g.][]{oppenheimer20,chadayammuri22}. Specifically, we utilized the pyXSIM package \citep{zuhone16}, which generates mock photons via a Monte Carlo sampling of X-ray spectra \citep{2012MNRAS.420.3545B} from the Astrophysical Plasma Emission Code (\textsc{apec}) from each simulation fluid element with $T > 10^{5.3}$~K and hydrogen number density $n_{\rm H} < 0.22 \ \rm{cm^{-3}}$. We considered fluid elements that belong to the galaxy halo and extend out to three virial radii. The convolution of the CGM emission with the LEM instrument model, the Galactic foreground emission and absorption, and the CXB sources were simulated using the SOXS software tool\footnote{\url{https://hea-www.cfa.harvard.edu/soxs}}. The spectrum of the CXB sources and the Milky Way foreground are described in Section \ref{sec:data}. We note that we generated multiple CXB realizations and each galaxy has a separate CXB image. However, the image representing the foreground emission is the same for all mock images. The mock projected observations have an exposure time of $10^6$~s and utilize the LEM pixel size ($15\arcsec$) and spatial resolution ($10\arcsec$ half-power diameter). The galaxies are placed at $z=0.035$. We note that the mock images do not include the emission from X-ray binaries, the interstellar medium, or any AGN emission. However, these X-ray-emitting components are confined to within the stellar body of the galaxy and would not contribute to the large-scale CGM emission that is the subject of the present study.

\begin{table*}
\caption{Characteristics of the simulated Magneticum galaxies}
\begin{minipage}{19cm}
\renewcommand{\arraystretch}{1.4}
\scriptsize	
\resizebox{0.95\textwidth}{!}{%
\begin{tabular}{ccccccccccccc}
\hline
ID &  $M_{\rm 200}$ & $M_{\rm \star}$  &  SFR &  \multicolumn{2}{c}{$R_{\rm 200}$} &  \multicolumn{2}{c}{Area} & $N_{\rm \ovii}$ &  $N_{\rm \oviii}$ &  $N_{\rm \fexvii}$ &  $N_{\rm \neix}$ &  $N_{\rm \nex}$  \\
           & ($\rm{10^{10} \ M_{\rm \odot}}$) & ($\rm{10^{10} \ M_{\rm \odot}}$) & ($\rm{M_{\rm \odot} \ yr^{-1}}$) & (kpc)& ($\arcmin$)& ($\rm{arcmin^2}$) & FOV & ($\rm{cm^{-2}}$) & ($\rm{cm^{-2}}$) & ($\rm{cm^{-2}}$) & ($\rm{cm^{-2}}$) & ($\rm{cm^{-2}}$) \\
(1) & (2) & (3) &(4) & (5) & (6) & (7) &(8) & (9) & (10) & (11) & (12) & (13) \\
\hline
12623 & $1286.83$ & $32.43$& $4.26$ & $590.10$& 14.32 &  $586.23$ & $0.651$ &$1.59 \times 10^{15}$ & $2.01 \times 10^{15}$ &  $8.67 \times 10^{13}$   &  $6.36 \times 10^{14}$  &  $2.08 \times 10^{14}$  \\
13252 & $928.69$ & $27.26$ & $7.93$ & $529.32$& 12.84 & $471.31$ & $0.524$ &$3.19 \times 10^{15}$ & $2.65 \times 10^{15}$ &  $4.65 \times 10^{13}$  &  $7.86 \times 10^{14}$  &  $1.61 \times 10^{14}$   \\
13570 & $993.79$ & $24.91$ &$2.13$ & $541.38$& 13.13 & $492.84$ & $0.548$ & $1.61 \times 10^{15}$ & $1.35 \times 10^{15}$ &  $1.21 \times 10^{13}$  &  $3.51 \times 10^{14}$ &  $9.13 \times 10^{13}$  \\
13633 & $939.97$ & $28.56$ &$8.44$ & $531.43$& 12.89 &$474.99$  & $0.528$ & $2.34 \times 10^{15}$ & $1.85 \times 10^{15}$ &  $2.98 \times 10^{13}$  &  $5.95 \times 10^{14}$ &  $1.08 \times 10^{14}$  \\
\hline
18178 & $256.53$ & $7.66$ & $1.61$ & $344.72$ & 8.36 & $652.63$  & $0.725$ & $8.16 \times 10^{14}$ & $9.90 \times 10^{14}$ &  $5.11 \times 10^{12}$   &  $1.80 \times 10^{14}$  &  $8.15 \times 10^{13}$ \\
18685 & $206.33$ & $8.30$ & $2.66$ & $320.58$ & 7.77 & $563.77$ & $0.626$ &$1.10 \times 10^{15}$ & $7.85 \times 10^{14}$ &  $5.90 \times 10^{11}$   &  $1.60 \times 10^{14}$  &  $4.72 \times 10^{13}$ \\
19665 & $168.43$ & $5.57$ & $2.17$ & $299.61$ & 7.27 & $493.55$ & $0.548$ & $5.86 \times 10^{14}$ & $4.71 \times 10^{14}$ &  $4.33 \times 10^{11}$   &  $6.48 \times 10^{13}$  &  $2.49 \times 10^{13}$ \\
19722 & $158.06$ & $6.64$ & $0.92$ & $293.33$ & 7.11 & $472.06$ & $0.525$ & $1.13 \times 10^{15}$ & $9.22 \times 10^{14}$ &  $3.21 \times 10^{12}$   &  $1.93 \times 10^{14}$  &  $7.01 \times 10^{13}$ \\
20115 & $126.75$ & $6.08$ & $14.79$ & $272.51$ & 6.61& $408.00$ & $0.453$ & $4.25 \times 10^{14}$ & $4.39 \times 10^{14}$ &  $1.78 \times 10^{12}$   &  $6.50 \times 10^{13}$  &  $3.76 \times 10^{13}$ \\
\hline
\end{tabular}}
\vspace{0.15in}
\end{minipage}
Columns are as follows. (1){\it Magneticum} galaxy ID; (2) Virial mass of the galaxy; (3) Stellar mass of the galaxy; (4) Star formation rate of the galayx; (5) and (6) The virial radius of the galaxy in kpc and in arcmin at z=0.035, respectively; (7) and (8)  The area of the studied annulus around galaxies expressed in $\rm{arcmin^2}$ and as a fraction of the $30\arcmin\times30\arcmin$ field-of-view, respectively. For massive galaxies the area is calculated using the $(0.3-1)R_{\rm 200}$ region while for Milky Way-type galaxies the $(0.3-1.75)R_{\rm 200}$ region is used; (9)-(13) Median column densities for the major ions.
\label{tab:galaxies}
\end{table*}

\subsection{Column density maps}
\label{sec:sims_description}

To simulate absorption in the outer parts of Milky Way-type and more massive galaxies, we used the redshift $z\sim 0$ outputs of the modern high-resolution cosmological hydrodynamical simulation {\it Box4/uhr} \citep{2015ApJ...812...29T} of the {\it Magneticum}\footnote{\url{www.magneticum.org}} simulation set \citep{2016MNRAS.463.1797D}.

This simulation is a box of (48 $h^{-1}$ cMpc)$^3$ comoving volume with dark matter m$_\mathrm{DM}=3.6\times 10^7 \ \rm{M_\odot}$ and gas m$_\mathrm{gas}=7.3\times 10^6 \ \rm{M_\odot}$ mass resolutions across it, with a total of 576$^3$ particles simulated with an improved version \citep{2016MNRAS.455.2110B} of the N-body code \textsc{Gadget~3}, which is an updated version of the code \textsc{Gadget~2} \citep{2005MNRAS.364.1105S} including a Lagrangian method for solving smoothed particle hydrodynamics (SPH). For the details on the simulation ingredients and resulting properties of the galaxy populations, we refer to \cite{2015ApJ...812...29T} and a recent analysis in \cite{2022arXiv220804975V}. The {\it Magneticum} simulations reproduce several key observational measurements that have been thoroughly explored and extensively described in previous studies \citep[e.g.][]{2015ApJ...812...29T,dolag17}. While part of this consistency originates from the initial calibration of various simulation parameters with the observational results, others are obtained independently of that. Notably, the scaling properties of the total X-ray emission as a function of galaxy stellar mass and star formation rate are fully consistent with the observed relations \citep{2023A&A...669A..34V}.

Importantly, the simulation follows the detailed evolution of various metal species and their relative composition in the circum- and intergalactic medium via computing continuous enrichment by supernovae of type Ia and type II, and asymptotic giant branch star winds in the fully cosmological context. The enrichment is computed self-consistently with the underlying evolution of the stellar populations \citep[for details, see][]{2007MNRAS.382.1050T} and matter outflows from the galaxies are triggered by star formation and AGN feedback \citep{2014MNRAS.442.2304H}. As a result, the simulations were capable of reproducing detailed metal distributions within galaxy clusters at redshift $z\sim 0$ \citep{2017Galax...5...35D}, while achieving already significant enrichment of the intergalactic medium already at $z\sim 2-3$ \citep{2017MNRAS.468..531B}. 

To build a representative set of massive disk galaxies, we used galaxies with virial masses comparable to the Milky Way and above, selected from  the poster-child disk galaxies sample of \citet{2015ApJ...812...29T}, selected according to their kinematic and morphological properties. We note that the selected galaxies do not reside in rich galaxy cluster or group environments. Such galaxies from this simulation have been shown to successfully reproduce many observational results.  Especially, detailed properties of galaxies of different morphologies can be recovered and well match observational findings. For example their angular momentum properties and the evolution of the stellar mass--angular momentum relation with redshift \citep{2015ApJ...812...29T,2016ilgp.confE..41T}, stellar kinematics of early-type galaxies \citep{2018MNRAS.480.4636S,2020MNRAS.493.3778S} and the size-mass relations and their evolution \citep[see e.g.][]{2016ilgp.confE..43R,2017MNRAS.464.3742R}. Also global properties like the fundamental plane \citep{2016ilgp.confE..43R}, the baryonic Tully-Fischer relation \citep{2022MNRAS.tmp.2826M}, the baryon conversion efficiency \citep[see e.g.][]{2015MNRAS.448.1504S,2017MNRAS.472.4769T}, as well as chemical properties \citep{2017Galax...5...35D,2021ApJ...910...87K} are in reasonable agreement with current observations. In addition, internal properties as dark matter fractions \citep{2017MNRAS.464.3742R}, in-situ and ex-situ fractions \citep{2022ApJ...935...37R} or the observed decline of rotation curves at high redshift are well reproduced \citep{2018ApJ...854L..28T}.

For each galaxy, a snapshot cutout a few cMpc in size has been extracted from the full simulation box, fully encompassing at least $\sim 2 R_\mathrm{vir}$ of the galaxies and keeping a portion of their direct IGM environment. The particle-based data of the cutout snapshots were first converted into the grid boxes of the physical quantities, namely gas density, temperature, metallicity, and three components of the velocity field. In all cases, the resulting values on the grid represent mass-weighted averages of the physical quantities.
 
 For the gas metallicity, only the single value giving the ratio of the metal mass to the corresponding metal mass for the Solar abundance of elements has been used, disregarding possible element-to-element variations in the partial abundances of the individual elements. For the CGM and dense IGM gas of interest here, only slight variations between oxygen and iron are visible in the {\it Magneticum} simulation, so this approach is fully justified, given the overall uncertainties in the treatment of production and spread of the metals in cosmological simulations.

Given the cubes of the physical parameters, the cubes of ion number densities have been calculated using the approach described in \cite{2001MNRAS.323...93C}. Namely, the gas is assumed to be in ionization equilibrium with respect to collisional and photoionization processes, with the latter driven by uniform Cosmic X-ray and UV radiation field of the double power law shape. In \citet{2019MNRAS.482.4972K}, the consistency of the outputs of this code with the calculations with Cloudy \citep{ferland17} and \citet{2012ApJ...746..125H} radiation field has been demonstrated. Linear scaling with metallicity was applied, neglecting any possible modifications of the equilibrium ionization state due to changes in the chemical composition of the gas.

As a result of these calculations, one can compute full Position-Position-Velocity cubes for any ion of interest, e.g.\ helium- and hydrogen-like oxygen, neon, or silicon, as well as L-shell ions of iron. For the latter, significant uncertainty in the predictions for the low-density photoionized plasma still exists, so one needs to allow for the possibility of substantial disagreement between different codes for some of the lines considered here (e.g.\ the \fexvii line at 826 eV). After the projection of the ion density cubes along any of the coordinate directions, one obtains the column density maps and column density velocity profiles for any region of interest. The column density maps of major X-ray ions of massive and Milky Way-type galaxies are presented in Figures \ref{fig:nhmaps_massive} and  \ref{fig:nhmaps_lowmass}.

\subsection{Galaxy selection}
\label{sec:galaxies}

To probe whether an X-ray IFU can detect absorption lines from the outer CGM, we probed galaxies with a range of dark matter halo masses. We split our investigation into massive galaxies and Milky Way-type galaxies. All selected galaxies are star-forming with $\rm{SFR} = 0.9-14.8 \ \rm{M_{\odot} \ yr^{-1}}$. The sample of four massive galaxies has a virial mass of $M_{\rm 200} = (0.9-1.3)\times10^{13} \ \rm{M_{\odot}}$ with a corresponding virial radius of $r_{\rm 200} =530-590$~kpc. Given that the galaxies are placed at  a redshift of $z=0.035$, the corresponding angular scale is $r_{\rm 200} =12.9\arcmin - 14.3\arcmin$,  implying that these galaxies fill the field of view. 

The Milky Way-type galaxies have viral mass of $M_{\rm 200} = (1.3-2.6)\times10^{12} \ \rm{M_{\odot}}$ and their virial radius is in the range of $r_{\rm 200} = 270-345$~kpc. Given the $30\arcmin\times30\arcmin$ field of view and the virial radius of these galaxies, we can study the CGM even beyond the virial radius. Therefore, the Milky Way-type galaxies are probed out to $1.75R_{\rm 200}$. The physical properties of the simulated galaxies are listed in Table \ref{tab:galaxies}.

\begin{figure*}
  \begin{center}
      \includegraphics[width=0.95\textwidth]{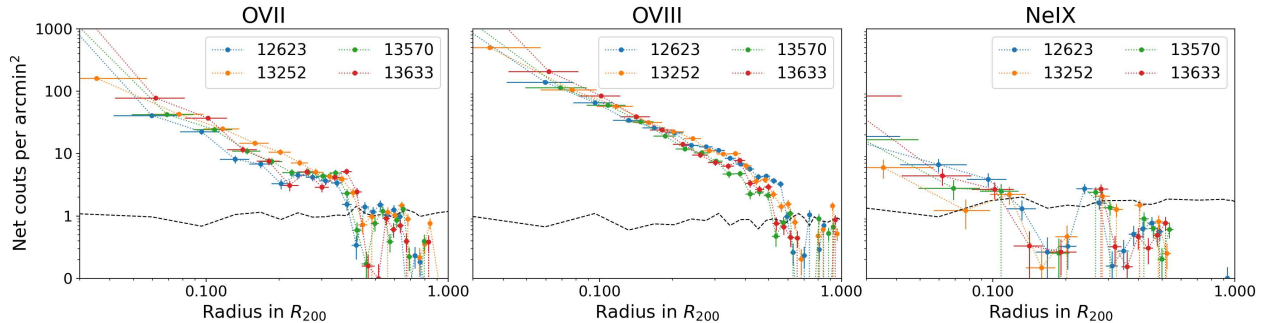}
      \caption{Surface brightness profile of four massive {\it Magneticum} galaxies extracted at the energy of \ovii (left), \oviii (middle), and \neix (right) emission lines. The profiles were extracted using a 2~eV wide bin around the energy of the redshifted emission lines. The dashed line represents $20\%$ of the combined level of the Milky Way foreground emission and the unresolved fraction of the CXB sources. The profiles have detectable emission to about $(0.4-0.6)R_{\rm 200}$ radius for the \ovii and \oviii emission lines.} 
     \label{fig:profiles}
  \end{center}
\end{figure*}

\subsection{The need for absorption studies}
\label{sec:why}

While the focus of the present work is to demonstrate that the CGM can be studied with absorption studies, we note that absorption studies are the only viable method to probe the CGM on the largest scales.  Figure \ref{fig:profiles} presents the azimuthally averaged \ovii (left), \oviii (middle), and \neix (right) surface brightness profiles of the four massive {\it Magneticum} galaxies in our sample. The analyzed mock event files include all main emission components, namely the CGM emission from the galaxies, the emission from the Milky Way foreground, and the CXB sources. To generate the surface brightness profiles, we excluded the  brightest 50 CXB sources  to decrease the contribution from the (resolved) CXB emission. To obtain the azimuthal surface brightness profiles, we extracted $2$~eV wide bins around the redshifted emission lines in azimuthal annuli that are centered on the galaxy. For the \ovii and \neix~profiles, we co-added the emission signal from the forbidden and resonance lines to increase the signal-to-noise ratios. In the figure, we also show the $20\%$ level of the total background emission that has already been subtracted from the profile. This level provides a conservative estimate of the accuracy of background subtraction. The surface brightness profiles and detectability of the CGM will be discussed in detail in an upcoming studies (Schellenberger et al. in prep; Truong et al. in prep). 

The surface brightness profiles reveal that the CGM around these massive galaxies can be traced to about $(0.4-0.6)R_{\rm 200}$ at the energy of the \ovii and \oviii emission lines. For the \neix lines, the emission is significantly fainter and more concentrated in the central parts of the galaxy and the emission from the \neix line is only detectable within $\sim0.1R_{\rm 200}$. As a consequence of the rather strong emission lines within $\sim0.3R_{\rm 200}$, it is not viable to study the CGM in absorption as the strong emission lines will dominate over the absorption lines. However, beyond this region the emission lines become progressively weaker, allowing us to study the CGM properties in absorption. Overall, we conclude that emission studies cannot probe the CGM at the largest scales ($\gtrsim0.5R_{\rm 200}$) and this extremely tenuous gas is only accessible by absorption studies.

\section{Generating the X-ray spectra}
\label{sec:data}

Based on the column density maps presented in the previous section, we simulated realistic X-ray spectra that include the absorption component from the CGM. These spectra were used to probe whether the absorption lines imprinted on the cumulative spectrum of CXB sources can be detected. To construct the X-ray spectrum, we included the following components: (1) emission from the brightest 50 CXB sources; (2) the Milky Way foreground emission enclosed within the source regions; (3) the diffuse emission associated with the galaxy at the location of the sources; and (4) the X-ray absorption lines with equivalent width inferred from the column density maps. In the next paragraphs, we discuss each of these components in detail. We do not include an instrumental background component in the model as its level is expected to be negligible compared to the CXB emission.  

The bulk of the X-ray emission in the simulated spectra originates from the population of CXB sources. To describe the X-ray spectrum of the bright AGN, we used a featureless power-law spectrum with a slope of $\Gamma = 1.47$, which is the typical spectrum of these sources \citep{hickox06}. To maximize the signal-to-noise ratios, we only utilize the 50 brightest sources in the simulated field of view. This choice represents a balance between maximizing the flux from the population of CXB sources while minimizing the emission from the Milky Way foreground. Specifically, the 50 brightest sources in a $30\arcmin \times 30\arcmin$ field represent more than $50\%$ of the total flux associated with all CXB sources, while these sources cover less than $2\%$ of the full field-of-view assuming the $15\arcsec$ pixel size. Because each galaxy has a different realization of CXB sources, we considered each field separately.  Accordingly, we carried out the source detection on each image and computed the corresponding emission from the CXB sources. We note that not all of the 50 brightest sources fall within the studied annuli around galaxies. Therefore, we only considered the emission from those point sources that are within the selected regions, i.e.\ $(0.3-1)R_{\rm 200}$ for massive galaxies and $(0.3-1.75)R_{\rm 200}$ for Milky Way-like galaxies. 

To describe the emission from the Milky Way foreground, we applied two main components, the emission from the Local Hot Bubble and the emission from the Hot Halo \citep{mccammon02}.  The Local Hot Bubble is modeled with an unabsorbed optically-thin thermal plasma (\textsc{apec}) model with $kT = 0.099$~keV. The Hot Halo component is described with two absorbed \textsc{apec} components with $kT = 0.225$~keV and $kT=0.7$~keV and the column density of $N_{\rm H} = 1.8 \times10^{20} \ \rm{cm^{-2}}$ was applied. The existence of the hotter component of the Hot Halo model is suggested by HaloSat and \textit{eROSITA} observations. The abundance of all thermal components was set to Solar. The normalizations of the two cooler \textsc{apec} models were taken from \citet{mccammon02}, while the normalization of the $kT=0.7$~keV component was set to $10\%$ of the $kT = 0.225$~keV component \citep{ponti22}. The emission associated with these components was normalized for each galaxy individually and this normalization was derived according to the combined area of the sources within the source extraction region. The 50 brightest point sources cover $1.7\%-2.2\%$  of the total field of view, while the sources that fall within the studied annuli fill $\sim1\%$ of the field of view. 

While in this work we focus on studying the X-ray emission in absorption, the galaxies may have a non-negligible X-ray emission from their CGM at the position of the CXB sources. Because this component will produce an emission line at the same energy, the depth of the absorption line will decrease or may completely fade if the CGM emission is  bright. To account for the CGM emission from the galaxies, we extracted the X-ray spectrum of the diffuse gaseous component at the position of the CXB sources that fall within the studied region of the galaxies. We used the mock event files of the diffuse gas and fit the galaxy emission with an \textsc{apec} model. For the fitting, we allowed the temperature, abundance, and normalization to vary, but the redshift was fixed at $z=0.035$. For the four massive galaxies, we found that the best-fit gas temperatures were in the range of $kT=0.27-0.40$~keV. We used the model spectra obtained for the individual galaxies and added this to the overall spectral model. For the Milky Way-type galaxies, the CGM emission does not play a significant role and was therefore neglected. 

\begin{figure*}[!htp]
  \begin{center}
      \includegraphics[width=0.95\textwidth]{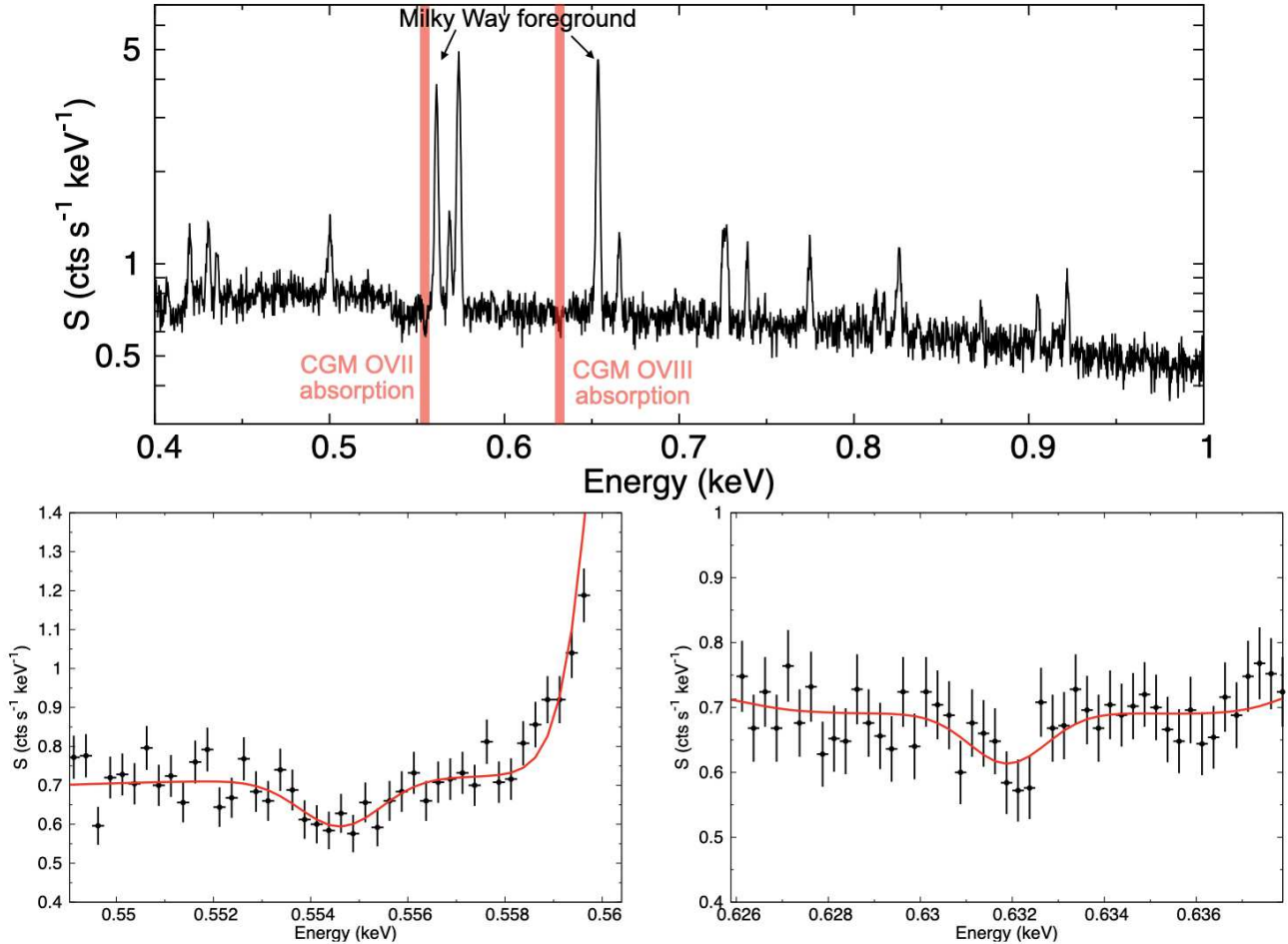}
      \caption{An example simulated spectrum of the CXB sources in the $(0.3-1)R_{\rm 200}$ region of the {\it Magneticum} galaxy \#13252, which is the second most massive galaxy in our sample with $M_{\rm 200} = 9.3\times10^{12} \ \rm{M_{\odot}}$. All main X-ray emitting components are included in the spectrum, including the emission from CXB sources, the Milky Way foreground emission at the location of the point sources, the emission from the large-scale CGM, and the absorption lines. The top panel shows the $0.4-1$~keV band spectrum. The emission lines originate from the Milky Way foreground emission while the continuum level is mostly determined by the CXB emission. The shaded regions mark the position of the \ovii and \oviii absorption lines. The bottom panels show the same spectrum, but in narrow energy ranges around the \ovii (left) and \oviii (right) absorption lines. The best-fit model is shown with the red curves. For this particular realization, the detection significances are $6\sigma$ for the \ovii and $3.5\sigma$ for the \oviii absorption lines.} 
     \label{fig:spectrum}
  \end{center}
\end{figure*}

The final components of the model are the absorption lines, which were modeled with a negative Gaussian line (\textsc{agauss} in \textsc{XSpec}). The normalization of the Gaussian lines was set to match the median column density within the $(0.3-1)R_{\rm 200}$ region for massive galaxies. The median column density within the annuli was derived from the median pixel value of the column density maps within the $(0.3-1)R_{\rm 200}$ annulus. To derive the equivalent widths of the lines from the column density, we assumed that the source is optically thin and, hence, the absorption line is unsaturated. 

The final model includes the sum of the above-described components, all of which were derived individually for each simulated galaxy. To obtain statistically meaningful conclusions, we generated 1000 simulations for each galaxy in our sample using the \textsc{fakeit} command in \textsc{xspec}. These data were used to compute the detection significance of the absorption lines. We present histograms of the detection significance in the next section.  

\begin{figure*}
  \begin{center}
      \includegraphics[width=0.48\textwidth]{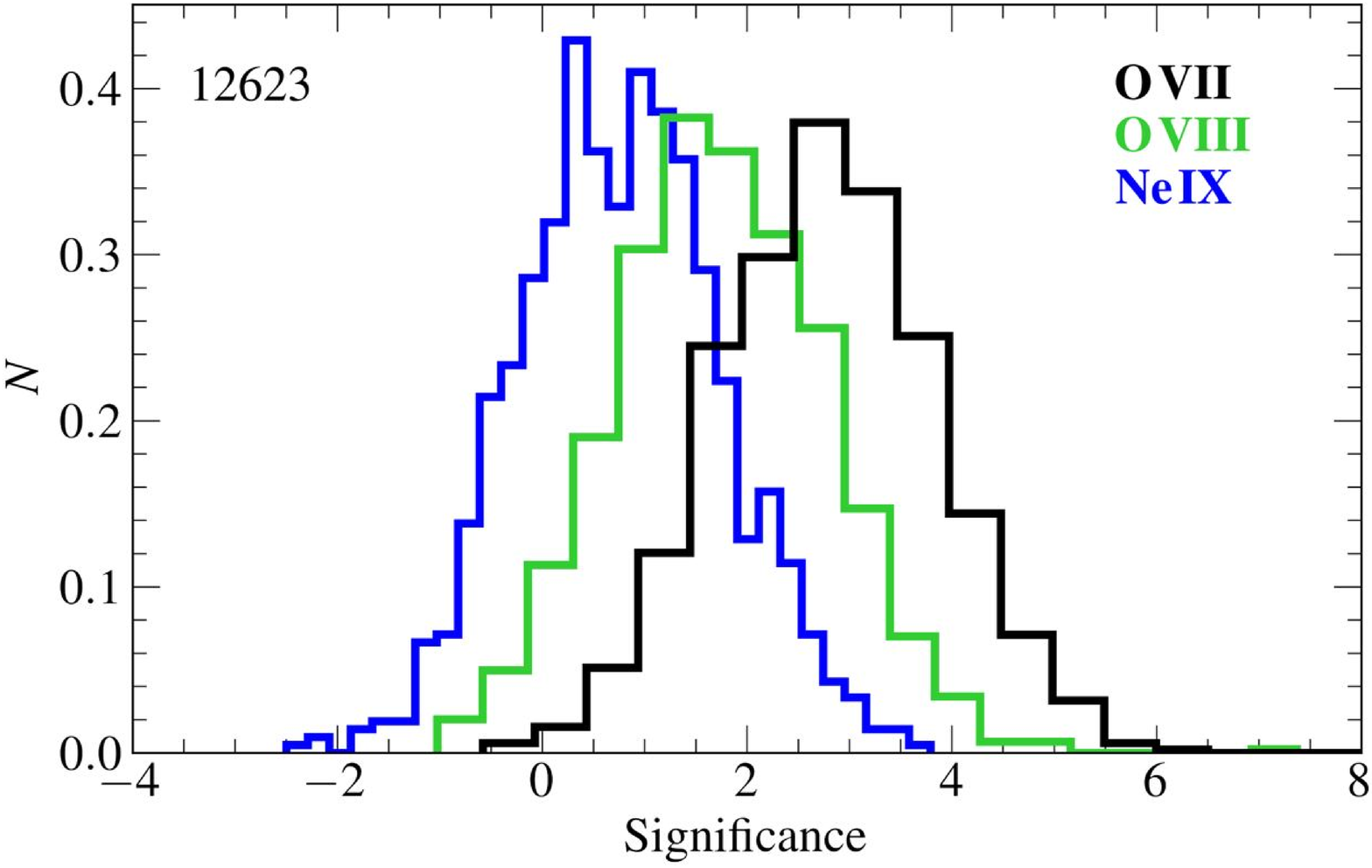}
      \includegraphics[width=0.48\textwidth]{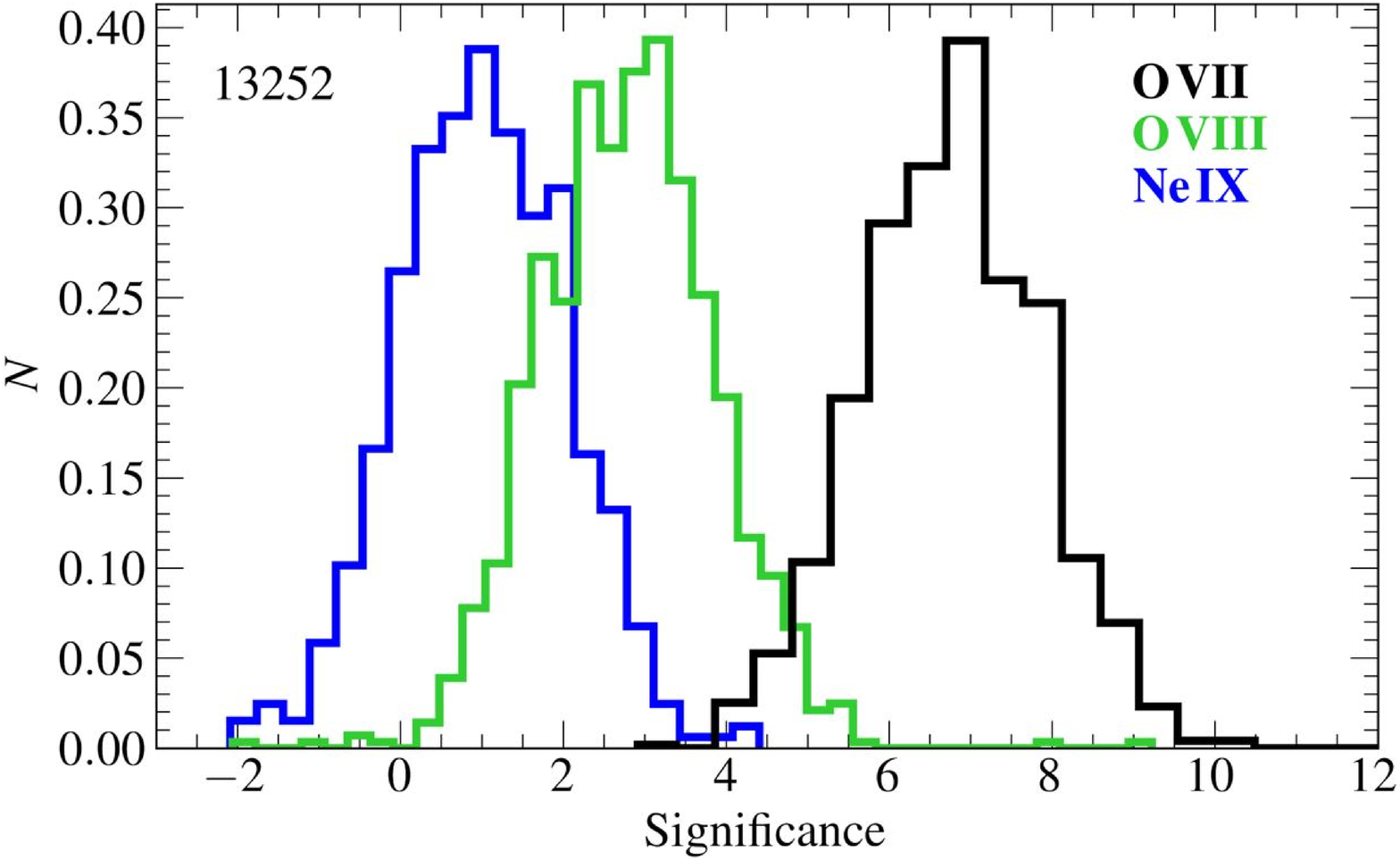}
      \includegraphics[width=0.48\textwidth]{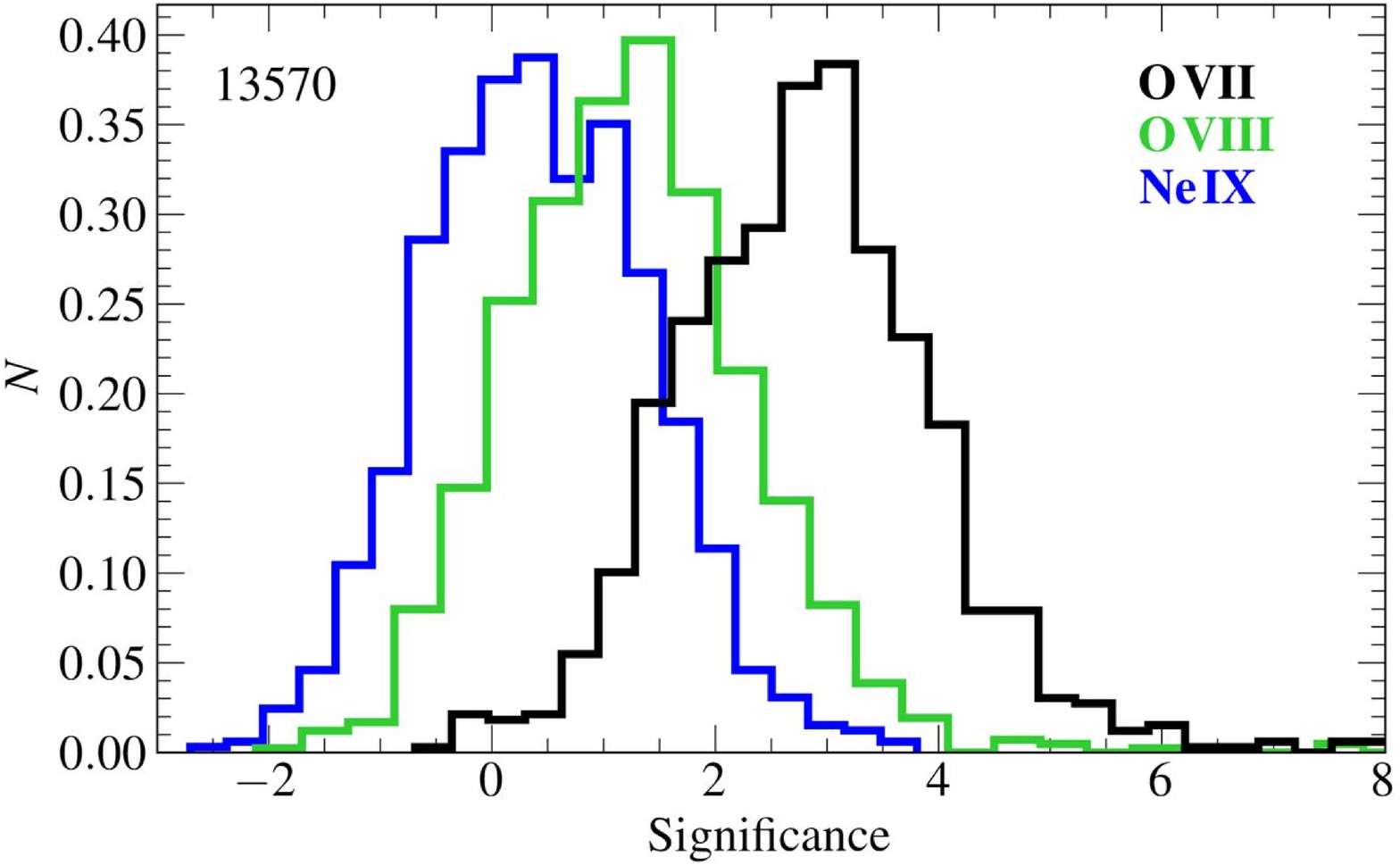}
      \includegraphics[width=0.48\textwidth]{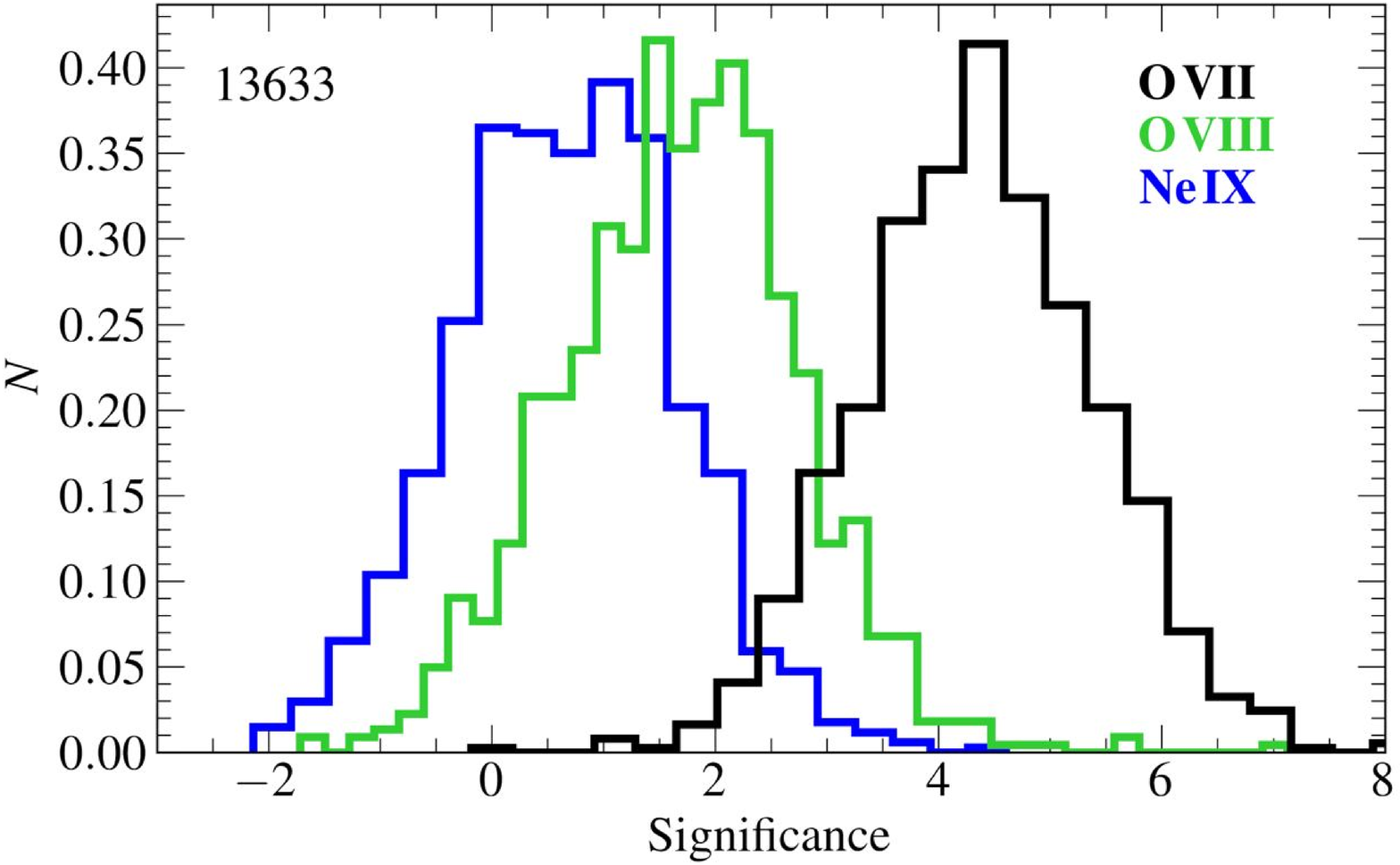}
      \caption{Simulated detection significances of the strongest absorption lines for the four massive {\it Magneticum} galaxies. To create the histograms, 1000 \textsc{XSpec} simulations were carried out for each galaxy. The most significant detection is expected from the \ovii absorption line and weaker detection is predicted from the \oviii absorption lines. The \neix line is not expected to be detected in $10^6$~s deep exposures with the notional instrument parameters.} 
     \label{fig:significances_massive}
  \end{center}
\end{figure*}

\section{Results}
\label{sec:results}

\subsection{Fitting the simulated spectra}
\label{sec:fits}

The simulated spectra were fit with a model that included all the components introduced in the previous section. Specifically, we included the emission components of the Milky Way foreground, the CXB emission, and Gaussian absorption lines. For the fits, the energy of the absorption lines was fixed, but their normalization was allowed to vary. For each simulated spectrum, we computed the normalization and the $1\sigma$ uncertainties of each absorption line. This procedure was repeated for all simulated spectra. 

In the top panel of Figure \ref{fig:spectrum} we present an example simulated spectrum for the massive {\it Magneticum} galaxy \#13252. The upper panel shows the $0.4-1$~keV band spectrum that includes all components described above. The visual appearance of the spectrum is largely determined by the Milky Way foreground emission lines, most notably the \ovii line complex at $E=561-574$~eV and the strong \oviii line at $E=654$~eV. The overall continuum emission is determined by the emission from the ensemble of CXB sources. The top panel of  Figure \ref{fig:spectrum} also highlights the location of the redshifted \ovii and \oviii absorption lines at energies $E=554.5$~eV and $E=631.9$~eV, respectively. While the absorption lines are identifiable in the broad band spectrum, we present narrow-band spectra and the best-fit models of these regions in the bottom panels of Figure \ref{fig:spectrum}. For this particular spectrum, which was obtained from a $10^6$ s pointing of a simulated galaxy, we obtained a $\sim6\sigma$ detection of the \ovii absorption line (bottom left panel) and a $\sim3.5\sigma$ detection of the \oviii absorption line (bottom right panel), while the \neix line is only detected at the $\sim1\sigma$ level (not shown). The best-fit \ovii and \oviii column densities for this particular spectrum are $N_{\ovii} = 4.2\times10^{15} \ \rm{cm^{-2}} $ and $N_{\oviii} = 5.1\times10^{15} \ \rm{cm^{-2}} $, respectively.

\begin{figure*}
  \begin{center}
      \includegraphics[width=0.48\textwidth]{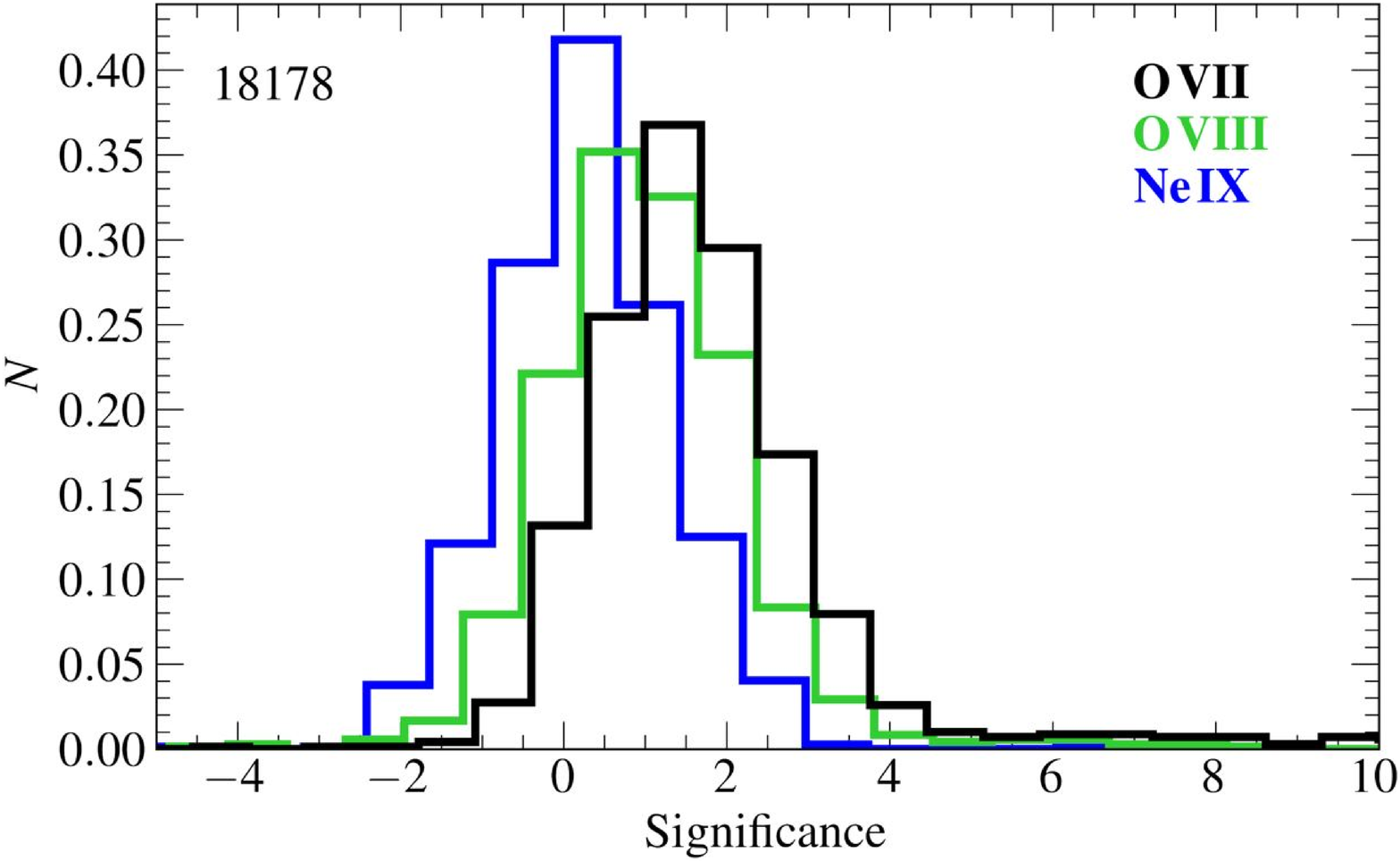}
      \includegraphics[width=0.48\textwidth]{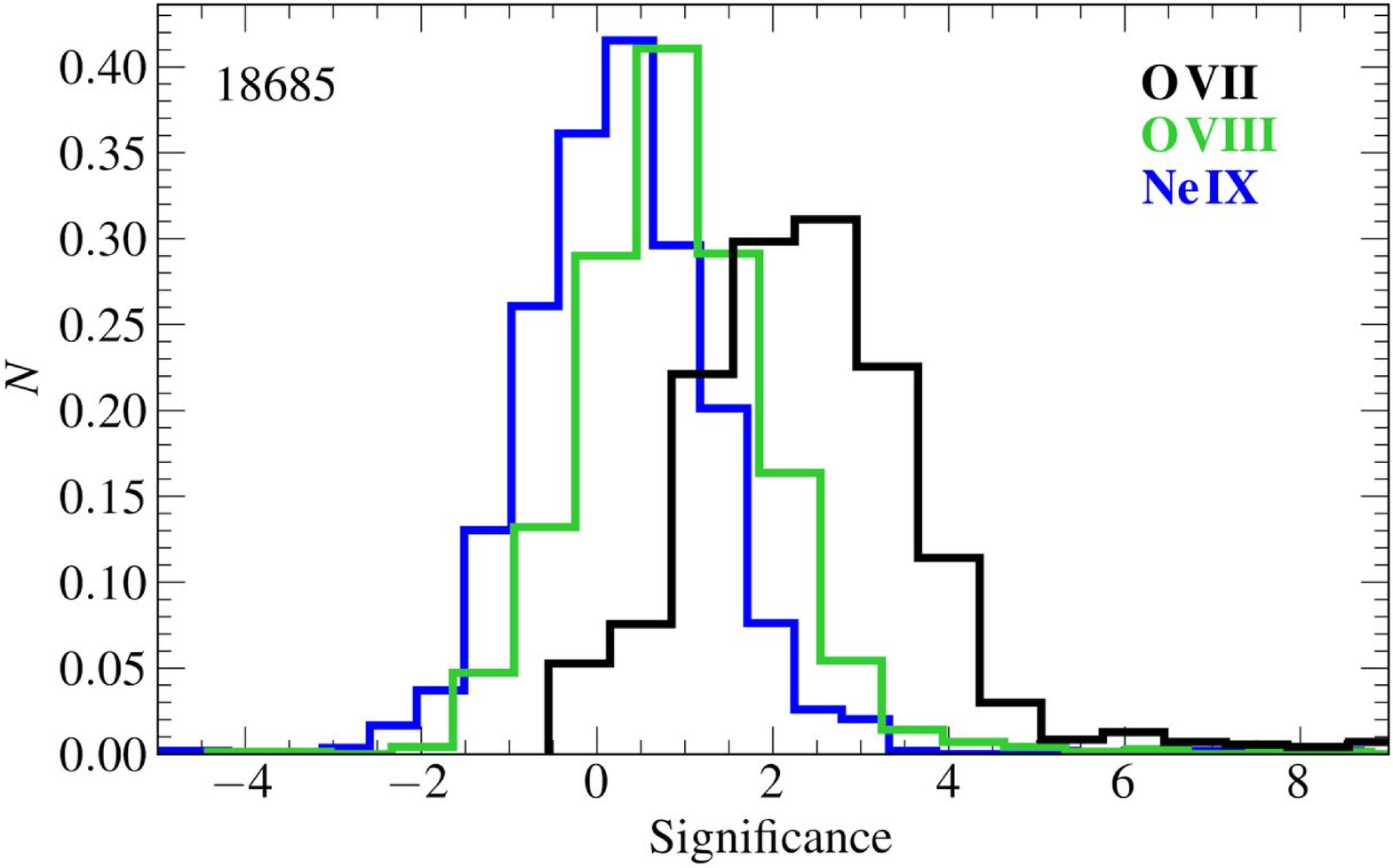}
      \includegraphics[width=0.48\textwidth]{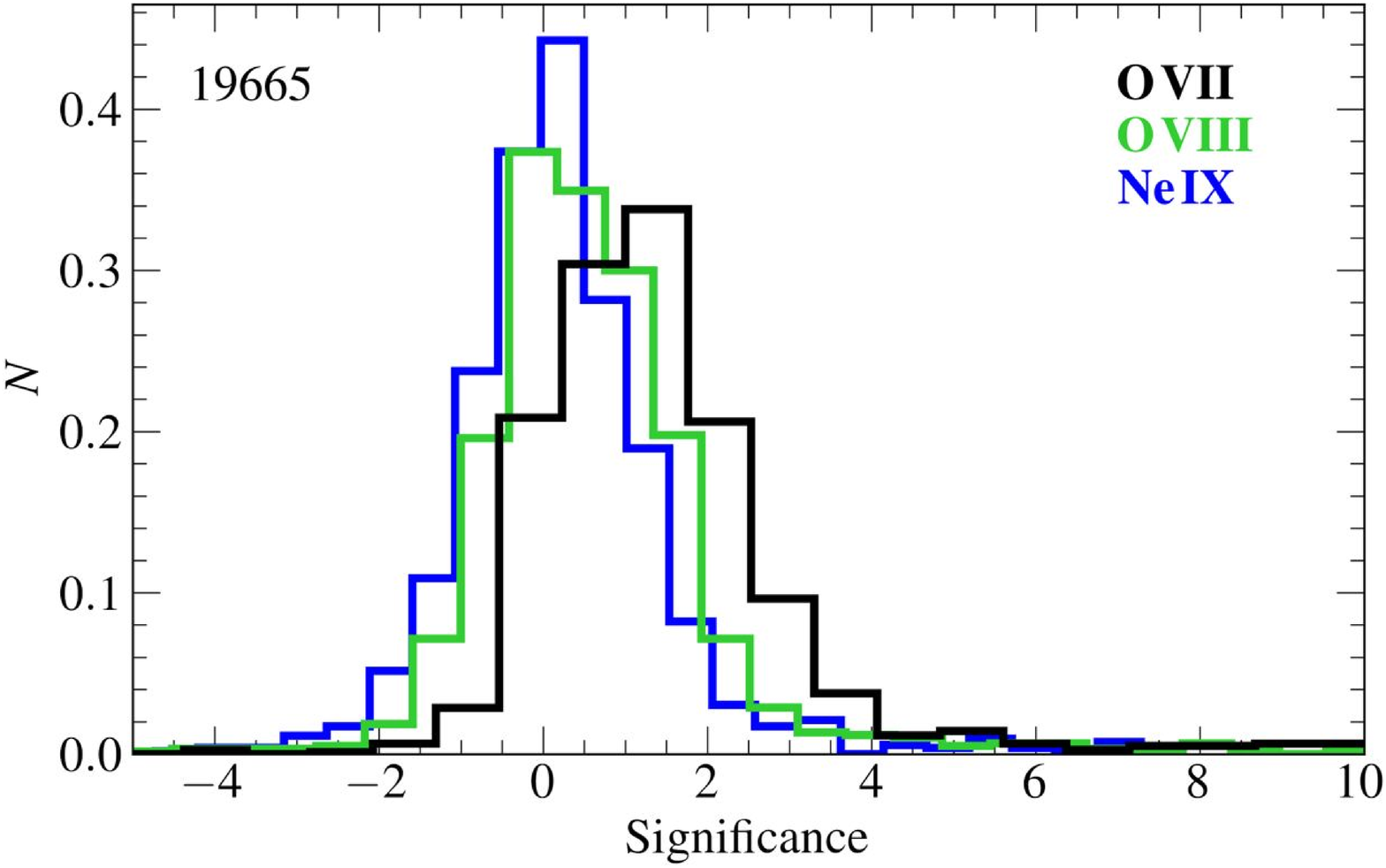}
      \includegraphics[width=0.48\textwidth]{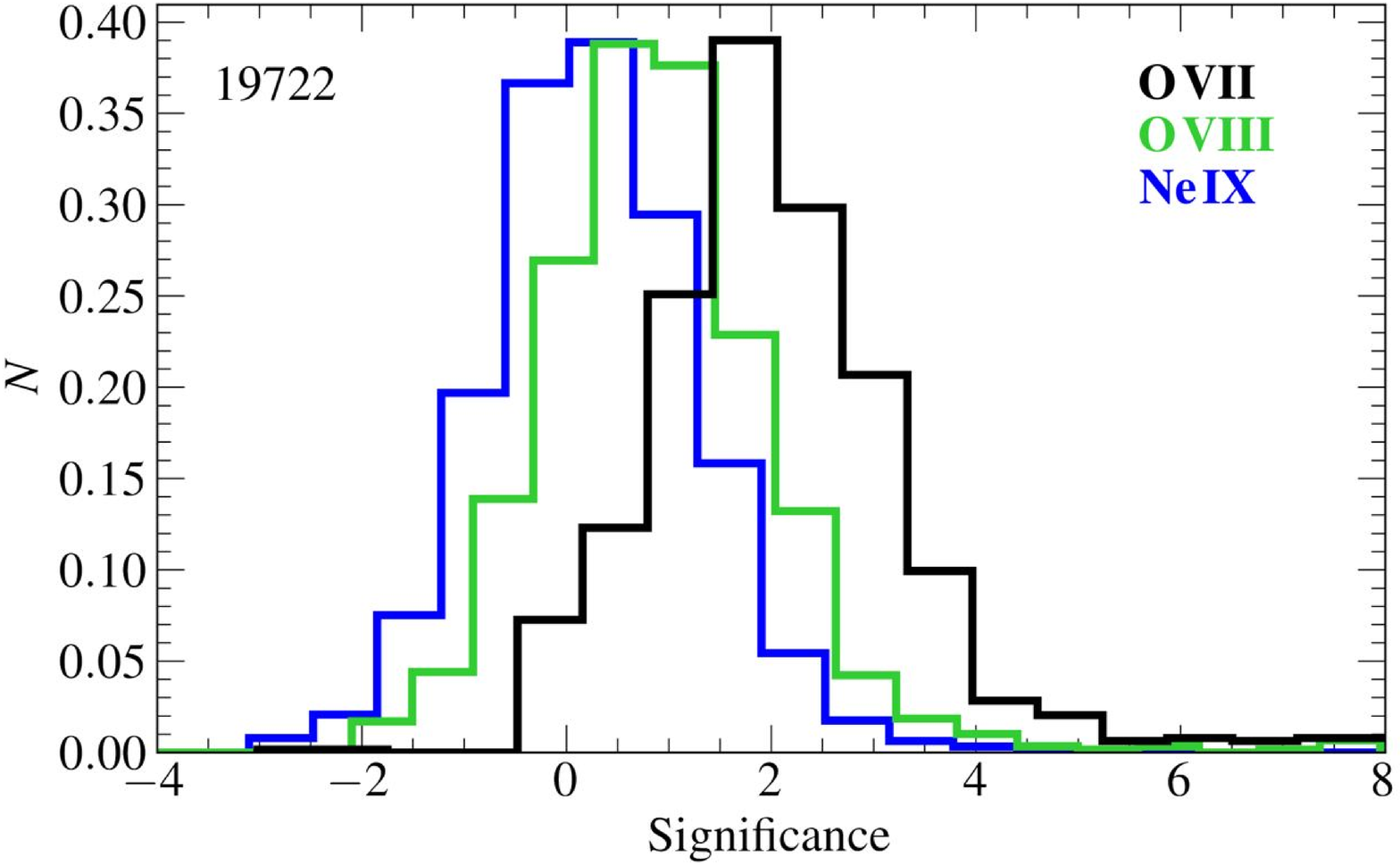}
      \includegraphics[width=0.48\textwidth]{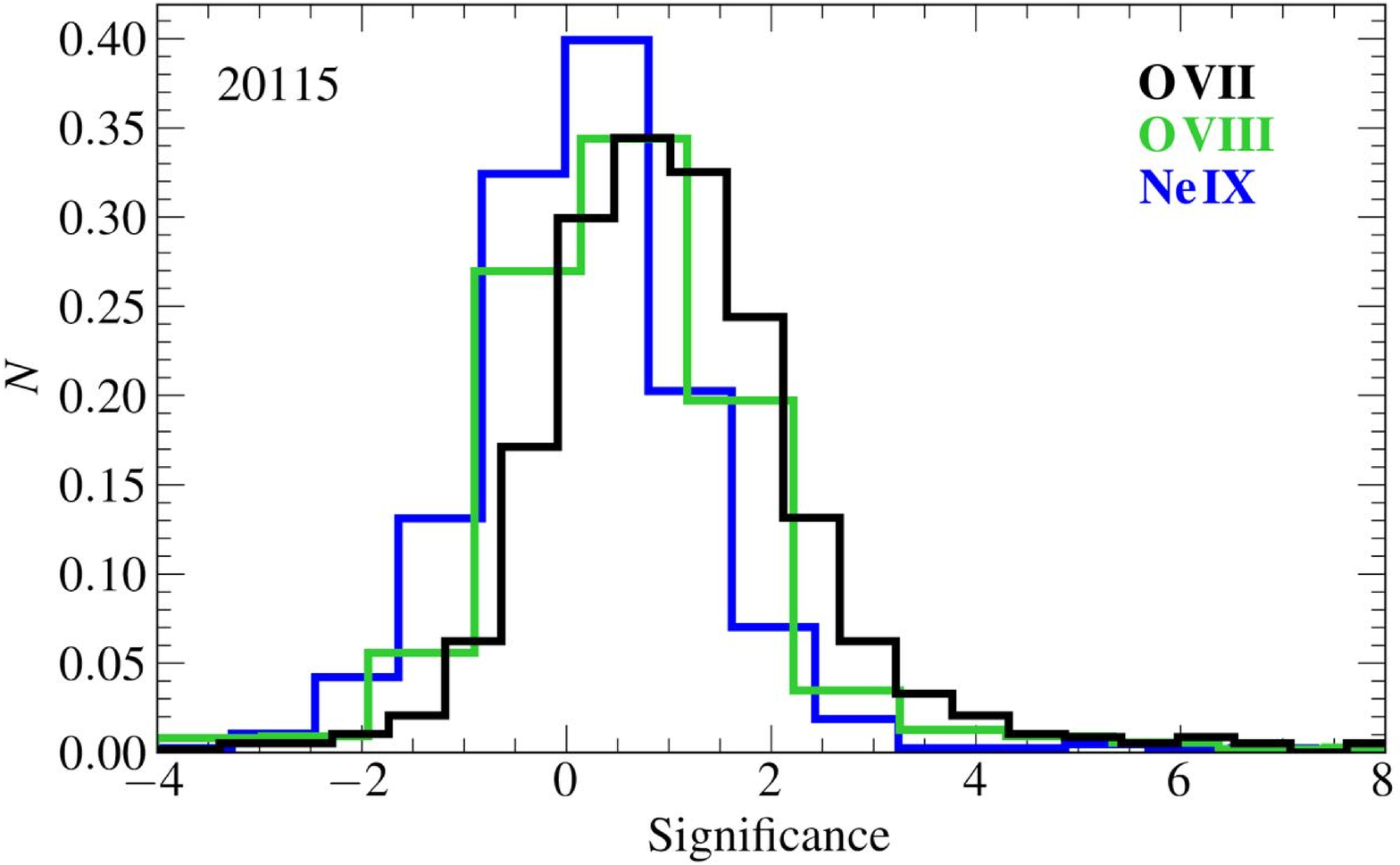}
      \caption{Same as Figure \ref{fig:significances_massive}, but for the five Milky Way-type {\it Magneticum} galaxies in our sample. The spectra were extracted for CXB sources that fall within the $(0.3-1.75)R_{\rm 200}$ region of the galaxies. For these lower mass galaxies, the \textsc{xspec} simulations predict weaker detections or non-detections of the \ovii absorption line and non-detections for the \oviii and \neix absorption lines.} 
     \label{fig:significances_mw}
  \end{center}
\end{figure*}

\subsection{CGM around massive galaxies}
\label{sec:massive_galaxies}

The histograms of detection significances presented in Figure \ref{fig:significances_massive} encapsulate the main results of the spectral fitting analysis for the four massive {\it Magneticum} galaxies. The most significant detection is from the \ovii line, which is detected with a median significance of $3\sigma-6\sigma$ for all massive galaxies. The median detection significances for the \oviii~lines are in the $1\sigma-3\sigma$ range, implying weak detections or non-detections. The weakest signal arises from the \neix line, which remains undetected in every galaxy. We do not present the detection significances for the other absorption lines listed in Table \ref{tab:galaxies}, which also remain undetected.  When investigating the best-fit normalizations of the Gaussian absorption lines, we find that the simulations well recover the median column densities of the simulations.  

Figure \ref{fig:significances_massive} demonstrates that the predicted detection significance of absorption lines exhibits strong galaxy-to-galaxy variation. The primary reason for this variation is caused by the different \ovii and \oviii column densities of the galaxies. Specifically, the highest detection significance is observed for galaxies \#13252 and \#13633, which also have the highest \ovii column densities (Table \ref{tab:galaxies}). As opposed to this, galaxies \#12623 and \#13633 have about a factor of two lower \ovii column densities yielding us lower detection significances. For the \oviii line a similar correlation can be established between the \oviii column density and the observer detection significances. 

We also probe whether the number of CXB sources may be responsible for some of the variations in the detection significances. We calculate the number of CXB sources and the total number of X-ray counts associated with these sources in the $(0.3-1)R_{\rm 200}$ region of the galaxies. We do not find a correlation between the CXB source counts and the detection significances. Specifically, the largest number of CXB sources and X-ray counts are associated with the outskirts of galaxy \#13633, while the fewest CXB sources reside in the $(0.3-1)R_{\rm 200}$ region of galaxy \#13252. We conclude that the primary parameter in determining the detection significances is the column density of the hot X-ray gas while the stochastic variability in the CXB realizations plays a lesser role. However, we note that the total flux from CXB sources is identical for all realizations, but the different realizations have stochastic variations.

We conclude that the large-scale CGM of massive galaxies can be robustly probed beyond $0.3R_{\rm 200}$ out to the virial radius using absorption lines using the cumulative spectrum of CXB sources. While the detection significances exhibit galaxy-to-galaxy variations, statistically significant detections at the $\gtrsim3\sigma$ level can be obtained for \ovii column densities of $N_{\rm \ovii} \gtrsim 1.5 \times 10^{15} \ \rm{cm^{-2}}$ in the  $(0.3-1)R_{\rm 200}$ region using $10^6$~s LEM observations. The detection of the \ovii line will allow constraining the basic characteristics of the outskirts of the CGM and additional detections of \oviii will allow estimating the mean gas temperature as well. Further discussion on the constraining power of the absorption lines is presented in Section \ref{sec:discussion}.

Given that we probed a broad annulus, we explore if the absorption signal is dominated by the inner, $(0.3-0.5)R_{\rm 200}$, or the outer, $(0.5-1)R_{\rm 200}$, regions of the annulus.  Given that we only use the 50 brightest CXB sources in the entire field of view, the $(0.3-0.5)R_{\rm 200}$ annulus only contains a relatively small fraction ($\lesssim5-20\%$) of the brightest sources that reside in the broad $(0.3-1)R_{\rm 200}$ annulus, which corresponds to $2-6$ sources. Thus, the bulk of the CXB sources is located within the $(0.5-1)R_{\rm 200}$ annulus. In addition, we emphasize that the column density of ions only shows a mild decrease with increasing radius, which is highlighted in Figure \ref{fig:nhmaps_massive}.  This is the consequence of two competing effects: while the gas density decreases with increasing radius, the integrated path length increases at larger radii, thereby compensating for the lower gas densities. The relatively weak dependence of column density on the radius is also highlighted in other simulations, such as EAGLE \citep{wijers20}.  With these considerations in mind, we repeated our stacking analysis using only the X-ray sources in the inner and outer regions. We conclude that the detection significances in the $(0.5-1)R_{\rm 200}$ annulus are virtually identical to those presented in Figure \ref{fig:significances_massive}, implying that using the cumulative signal from CXB sources probes the CGM in the outermost regions ($>0.5R_{\rm 200}$) of galaxies.

\begin{figure}
  \begin{center}
        \includegraphics[width=0.48\textwidth]{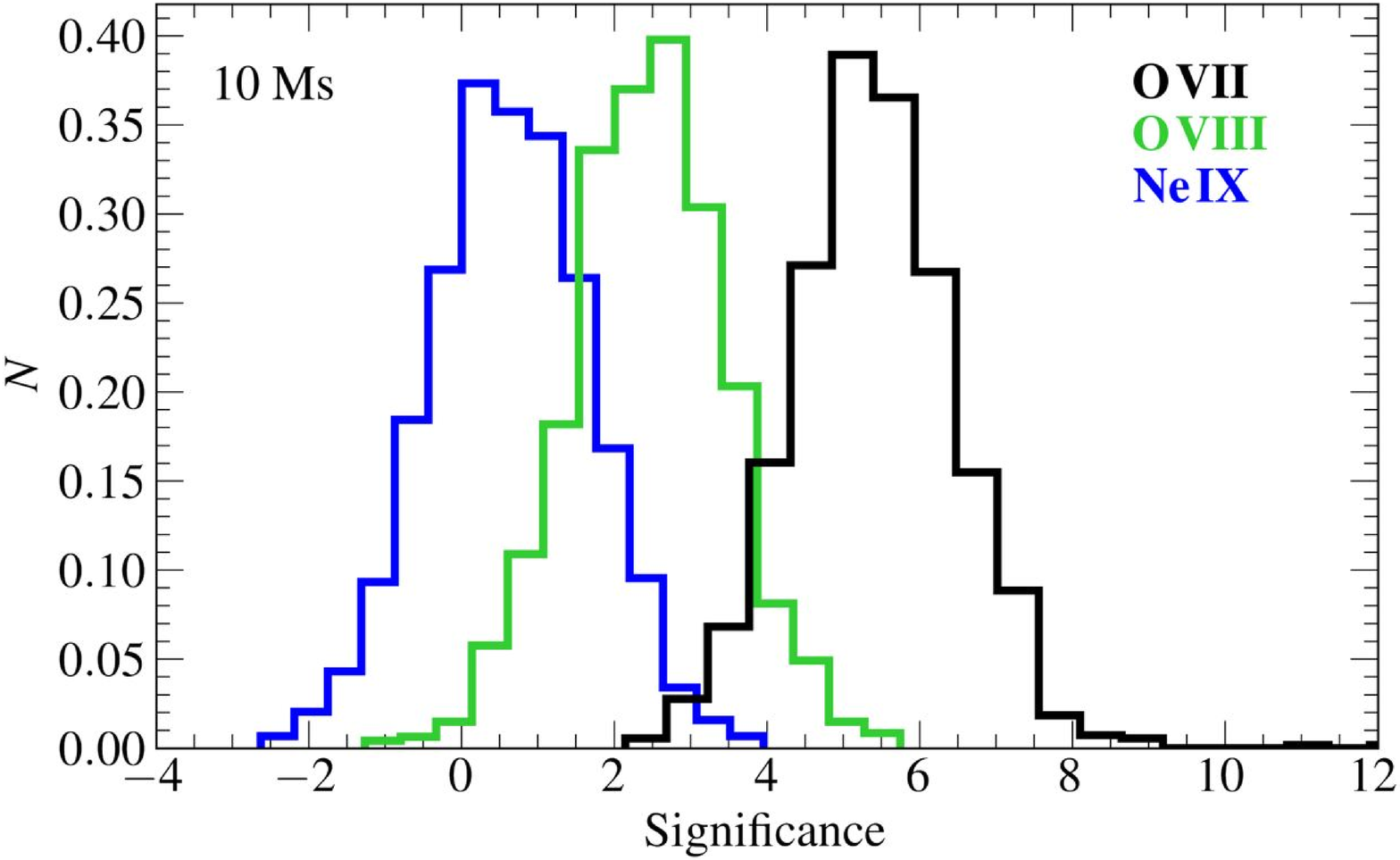}
      \caption{Simulated detection significances of stacked Milky Way-type galaxies with a total of $10$~Ms exposure time. To CXB sources were extracted from the $(0.3-1.75)R_{\rm 200}$ region around a galaxy with $300$~kpc virial radius. The stacked data will result in statistically highly significant, $\sim6\sigma$, detection of the \ovii absorption line and $\sim3\sigma$ detection of the \oviii absorption line in the cumulative spectrum of CXB sources. The \neix line will not be detected due to the  relatively low temperature of the large-scale CGM.} 
     \label{fig:significances_mw_stack}
  \end{center}
\end{figure}

\subsection{CGM around Milky Way-type galaxies}
\label{sec:mw_galaxies}

In the previous section, we demonstrated that studying the large-scale CGM of massive galaxies can reveal absorption lines. Here, we explore if the outskirts of Milky Way-type galaxies may also yield the detection of absorption lines. To this end, we probe the five star-forming galaxies with $M_{\rm 200} = (1.3-2.6)\times10^{12} \ \rm{M_{\odot}}$ (see Table \ref{tab:galaxies} for details), which is comparable with the Milky Way's virial mass. 

To probe whether absorption lines can be detected from the large-scale CGM, we explore the $(0.3-1.75)R_{\rm 200}$ region. While this region exceeds the virial radius of the galaxies, this allows us to probe the X-ray emitting gas at the largest scales and even consider the gas that may have been expelled from the galaxies beyond their virial radius. To generate the X-ray absorption spectra, we carried out a similar analysis as described in Section \ref{sec:data}. A major difference between the Milky Way-type and massive galaxies is that the lower mass galaxies do not contribute any significant X-ray emission beyond $0.3R_{\rm200}$ at the location of the point sources. Therefore, the emission component from the galaxy itself is negligible for this analysis. As before, we carried out 1000 \textsc{XSpec} simulations for each galaxy and summarize the results in Figure \ref{fig:significances_mw} that describe the distribution of detection significances for the five Milky Way-like galaxies. The plots suggest that the CGM of individual galaxies will either remain undetected (mostly the \oviii~and \neix~lines) and only the \ovii line galaxy may be weakly detected at the $2\sigma-3\sigma$ level. 

While the detection significances of major absorption lines remain below $3\sigma$ for individual Milky Way-type galaxies, we also explore whether the combined signal from these lower-mass galaxies can provide significant detection. Although this method does not inform us about the galaxy-to-galaxy variation of the CGM properties, it can provide crucial information about the average CGM properties of Milky Way-type galaxies. To probe the feasibility of this method, we derived the median column densities of the Milky Way-type galaxies listed in Table \ref{tab:galaxies} and assume a total exposure time of $10^7$~s for the ensemble of five galaxies. For the virial radius of the ``typical'' galaxy, we assumed $R_{\rm 200}=300$~kpc, which is approximately the mean of the five studied {\it Magneticum} Milky Way-type galaxies. For the list of CXB sources, we used the realization associated with galaxy \#19665. The \textsc{XSpec} analysis was identical to those for individual galaxies, except for the longer assumed exposure time. The resulting detection significances are presented in Figure \ref{fig:significances_mw_stack}. The detection significances for the ``stacked'' galaxies reveal that both the \ovii and \oviii will be detected significantly at the $\sim5\sigma$ and $\sim3\sigma$ level, respectively. The absorption signal from \neix remains undetected, which is due to the relatively low temperature of the CGM of these galaxies. 

Similar to the more massive galaxies, the absorption signal from Milky Way-like galaxies is not dominated by the inner regions ($<0.5R_{\rm 200}$). Indeed, only a few CXB sources are located in the $(0.3-0.5)R_{\rm 200}$ annulus with most sources residing at larger radius. Therefore, the observed absorption signal is determined by the CGM at $>0.5R_{\rm 200}$. Thus, using the cumulative signal from CXB sources and studying the CGM of Milky Way-type galaxies in absorption will  allow the study of the CGM on the largest scales.

\section{Discussion}
\label{sec:discussion}

\subsection{Measuring the gas parameters}
\label{sec:physics}

Detecting the large-scale CGM around galaxies offers a wide range of opportunities to characterize the physical parameters of the gas, to constrain the baryon content of galaxies, and to provide the much-needed input for models of structure formation. In this section, we briefly summarize how these observations will drastically improve our understanding of galaxy formation. Although, in this work, we focused on the detectability of X-ray absorption lines in single annuli around the galaxies, the analysis of CXB sources can be used to map properties of the galaxies, for example, by establishing gas density, gas mass, or temperature profiles. For example, for the stacked profile of Milky Way-type galaxies, it will be possible to establish the \ovii column density profile in $3$ bins while still having $\gtrsim3\sigma$ detection in each bin. 

The detection of absorption lines from multiple ions from the same chemical element allows the determination of the gas temperature. Given the expected gas temperature of the large-scale CGM, the transition strength, and the elemental abundances, the $f(\oviiin)/f(\oviin)$ ratios represent a diagnostics of gas temperature. Thus, by measuring this ratio, it is feasible to derive the {\it average} gas temperature all the way to the virial radius (and even beyond) of Milky Way-type and more massive galaxies. It is important to realize that this measurement is complementary to emission studies, which may not have the necessary signal-to-noise ratios beyond $\sim0.3R_{\rm200}$ to probe gas temperatures. Therefore, by combining emission and the proposed absorption studies, it will become feasible to construct the temperature map of Milky Way-type and more massive galaxies from their core to the virial radius. Probing the same halos in both emission and absorption is critical because it provides powerful constraints on biases in emission-derived gas properties. For example, detecting the CGM both in emission and absorption will allow us to probe the clumping factor of the CGM. 

Past observations hint that local galaxies are missing a substantial fraction of their baryons. For example, deep \textit{XMM-Newton} observations of massive disk galaxies suggest that at least half but possibly two-thirds of their baryons are not contained within their virial radius. This suggests that a fraction of the galaxy's baryons were ejected from the dark matter halos and may be found beyond the virial radius. However, these studies did not measure the baryon content of galaxies beyond $0.15R_{\rm 200}$ directly but used simple extrapolations. By carrying out the proposed absorption studies, the equivalent width, and, hence the column density of the hot gas can be inferred. By estimating the galaxy's virial radius (e.g.\ from its stellar mass), the total gas mass within the virial radius can be derived. Moreover, for sufficiently significant detections or by co-adding the data from multiple galaxies, a column density and gas mass profile can be inferred. This, in turn, can directly measure the fraction of baryons contained within the virial radius. Moreover, utilizing CXB sources beyond the virial radius will probe the baryon content up to $\sim2R_{\rm 200}$ and address whether the baryons are located within this region as suggested by SZ studies \citep{bregman22}.

\subsection{Advantages and caveats}
\label{sec:caveats}

Employing the cumulative spectrum of CXB sources to probe the CGM properties of galaxies offers a major advantage over traditional absorption studies, which rely on  single sightlines. Most notably, if the gas is not uniformly distributed, the traditional dispersive spectroscopy (e.g.\ grating instruments) will be biased high or low depending on the gas distribution and the location of the background source. As opposed to this, in the proposed method, a substantial number ($\gtrsim20$) of CXB sources  that randomly sample the large-scale CGM are used to obtain a representative picture of the CGM. A further advantage of using this method is that virtually any (nearby) galaxies can be probed, which allows the study of a comprehensive and well-understood galaxy population. As opposed to this, dispersive spectroscopy is only applicable to galaxies that are in the line of sight of bright background sources.  

A further advantage of our proposed method is that it is fully complementary to emission studies. That is an X-ray IFU can map the properties of the CGM in emission using the main emission lines (such as \ovii or \oviii) in the relatively bright regions of galaxies ($\lesssim(0.3-0.5)R_{\rm 200}$ for Milky Way-type galaxies). In these regions, the emission from these lines may either suppress or weaken the absorption signal. However, beyond these regions, the absorption lines in the cumulative spectrum of CXB sources will provide crucial information on the CGM using the same observations. Thus, IFU observations of galaxies can be simultaneously used for emission and absorption studies in the same pointing and carry out fully complementary analysis from the innermost regions of galaxies to their virial radius and beyond.

A potential limitation of absorption studies may be introduced due to the clumpiness of the large-scale CGM. While the level of clumpiness is not constrained by present-day observations, the limiting case is when all the X-ray CGM is in small dense clouds and the CXB sources are not coincident with any of these clouds. In this limiting case, the cumulative spectrum of CXB sources would not result in any absorption signal from the CGM. While the clumpy CGM may remain undetected in absorption, the dense, and hence, luminous X-ray gas clouds will be detected in emission. Therefore, the extreme clumpiness of the CGM may render studies with traditional dispersive spectroscopy unfeasible, while studies with an X-ray IFU remain fully viable.

Figures \ref{fig:nhmaps_massive} and \ref{fig:nhmaps_lowmass} reveal that the \textit{Magneticum} column density maps exhibit notable clumpiness. To assess the importance of clumping on the proposed absorption studies, we generated 1000 realizations of 25 randomly distributed CXB sources within the $(0.3-1)R_{\rm 200}$ annulus for each simulated massive galaxy in our sample. For each of the realizations, we derived the \ovii and \oviii detection significance following the method outlined in Section \ref{sec:results}. Given the clumpiness of the \textit{Magneticum} column density maps, none of the realizations reaches the limiting case where the CGM remains undetected in absorption. The randomly distributed sources recover the median column density values presented in Table \ref{tab:galaxies} and establish that the median column densities have a narrow distribution, which is comparable with the galaxy-to-galaxy variation. Thus, the clumpiness in the \textit{Magneticum} simulations does not affect the viability of the absorption studies.

A limitation of the present study is caused by the limited brightness of individual CXB sources, which, depending on the column density, requires co-adding the data from the $\gtrsim10$ brightest CXB sources. Therefore, our proposed method can only probe relatively large regions and detailed thermodynamic or column density maps cannot be constructed. To overcome these issues and study the CGM properties in narrower regions (e.g.\ annuli or sectors), the signal-to-noise ratios can be boosted by either deeper observations of individual galaxies or by co-adding the data from multiple galaxies with similar physical properties.  

Throughout this work, we assumed that the background AGN have a featureless spectrum that can be described with a power law model with a slope of $\Gamma = 1.47$. While this slope is representative as it was obtained for the ensemble of CXB sources in the Chandra Deep Fields \citep{hickox06}, the average spectrum of these sources was measured using \textit{Chandra} ACIS-I data, which has much coarser energy resolution ($130$~eV at $1.49$~keV energy) than an IFU instrument. Therefore, it is possible that a fraction of the CXB sources exhibit emission lines, which lines may not be detected at CCD resolution when co-adding the signal from a large number of CXB sources. Indeed, high-resolution spectroscopy of nearby Seyfert 2 galaxies demonstrates the existence of numerous emission lines originating from the circumnuclear emission region \citep{bogdan17,reeves17,kraemer20}. Depending on the redshift of the individual CXB sources and the ion species in the spectrum, it is feasible that some of the strong emission lines will be coincident with the expected wavelength of the absorption lines originating from the CGM. However, these emission lines are unlikely to make the analysis unfeasible since the spectrum of each CXB source can be investigated individually and if a strong redshifted emission line coincides with a wavelength of a CGM absorption line, the particular CXB source can be excluded from the analysis. While this will result in a small decrease in the signal-to-noise ratios, the overall analysis remains feasible. The exact effect of emission lines cannot be assessed as it depends on the redshift distribution of the sources and the emission lines present in the spectrum. However, given the high spectral resolution of the IFU instruments, line emission could be easily separated from the CGM emission, hence it is unlikely to play a significant role.

\subsection{Applicability of the method}
\label{sec:instruments}

To carry out the proposed analysis, a high-resolution non-dispersive X-ray instrument, i.e.\ an IFU is required. Traditional dispersive instruments, such as \textit{Chandra} LETG, \textit{XMM-Newton} RGS, or the proposed \textit{Arcus} mission with its grating spectrometer, can only study the spectrum of a single point source. Moreover, these instruments have (or are planned to have) much smaller effective area ($10-300\ \rm{cm^{2}}$ at $E=0.5$~keV), implying that only the brightest X-ray quasars can be effectively observed. Therefore, studying individual galaxies with these instruments is not feasible, and utilizing a multitude of (individually fainter) CXB sources to explore the large-scale CGM of nearby galaxies requires prohibitively long exposure times. 

The X-Ray Imaging and Spectroscopy Mission (\textit{XRISM}) will be launched in 2023, which will have an X-ray IFU that is primarily sensitive to X-rays above 2~keV energy with only $50\ \rm{cm^2}$ collecting area at $E=0.5$~keV energy. Moreover, the angular resolution of \textit{XRISM} is $75\arcsec$, implying that it will not be able to effectively differentiate the emission from CXB sources from the combined emission of the Milky Way foreground and the CGM emission. Additionally, \textit{XRISM} will have a small, $3\arcmin \times 3\arcmin$ field of view, hence it will not be able to cover the large apparent sizes of nearby galaxies. Thus, \textit{XRISM} will not be able to study the large-scale CGM of nearby galaxies. 
 
The X-ray Integral Field Unit (X-IFU) spectrometer will be the key instrument of the planned \textit{Athena} mission that will have $2.5$~eV spectral resolution. The collecting area of \textit{Athena} will be a factor of about 4 times larger than that assumed in this analysis ($\sim6000\ \rm{cm^2}$  at 0.5~keV energy), implying that the same signal-to-noise ratios can be achieved in shorter observations. Moreover, \textit{Athena} is designed to have  $5\arcsec$ spatial resolution, which would allow using smaller extraction regions for the CXB sources, thereby decreasing the contribution of both the emission from the Milky Way foreground and the emission from the extended CGM of the galaxy. However, the X-IFU instrument is planned to have a small, $5\arcmin \times 5 \arcmin$, which is not sufficient to cover nearby galaxies in a single pointing. For example, to cover the $(0.3-1)R_{\rm 200}$ region of the massive galaxies ($\sim500 \ \rm{arcmin^2}$) at $z=0.035$, about 20 pointing would be required. Because the larger collecting area and the better point spread function will not outweigh the effect of the small field of view, about $3$~Ms \textit{Athena} X-IFU observations would be required to map the CGM in the outskirts of a galaxy. Thus, the \textit{Athena} X-IFU would allow us to study the CGM in absorption, albeit studying the large-scale CGM in absorption would not be complementary to emission studies of the inner regions of the same galaxy.  

The ideal instrument to carry out the proposed analysis is a high spectral resolution non-dispersive spectrometer that has a large field of view, large collecting area, and moderate spatial resolution. The combination of these will allow the simultaneous mapping of the CGM in emission within the $\lesssim0.5R_{\rm 200}$ and probing the large-scale CGM to the virial radius and even beyond using absorption studies. These instrument parameters are synthesized in the proposed Line Emission Mapper (LEM) mission concept \citep{kraft22}. 

A key aspect of the proposed method is that it allows the study of the CGM around nearby galaxies in emission and absorption {\em simultaneously} using the same observations. Therefore, absorption studies relying on the population of CXB sources do not require additional exposures. Nonetheless, the method presented in this work is intended to complement traditional absorption studies rather than replace them. The brightest X-ray quasars have an X-ray flux of $10^{-12} - 10^{-11} \ \rm{erg \ s^{-1} \ cm^{-2}} $ in the $0.5-2$~keV band, which surpasses the X-ray flux of the 50 brightest CXB sources in the $30\arcmin \times 30\arcmin $ LEM field of view, which have a cumulative flux of about $6\times 10^{-13} \ \rm{erg \ s^{-1} \ cm^{-2}} $. Given the much brighter individual quasars, studying absorption lines imprinted on these bright background sources could, in principle, result in a higher detection significance. To this end, we carried out \textsc{XSpec} simulations assuming a bright X-ray quasar with a flux of $3\times 10^{-12} \ \rm{erg \ s^{-1} \ cm^{-2}} $ and assumed that the galaxy in its line of sight has a column density identical with those found for massive/Milky Way-type galaxies in our sample (Table \ref{tab:galaxies}). By generating 1000 realizations of 1~Ms deep LEM exposures and analyzing the data following Section \ref{sec:results}, the obtained median detection significance for the \ovii absorption line was $\sim\,8\sigma$ and $\sim\,3\sigma$ for the massive and Milky Way-type galaxies, respectively. These values exceed the detection significances attainable by relying on the cumulative spectrum of CXB sources (Section \ref{sec:results}). However, traditional absorption studies (1) are hindered by the limited galaxy population with the ideal redshift and stellar mass that lies in the line of sight of bright quasars, (2) are complicated by the fact that they only probe a pencil beam, which can mischaracterize the observed column densities (see Section \ref{sec:caveats}), and (3) require additional deep exposures and do not allow the simultaneous study of the same galaxies in emission. Thus, while probing absorption lines imprinted on the spectrum of bright background quasars remains a viable method to study the large-scale CGM, using the cumulative spectrum of CXB sources offers an exciting and complementary approach for the next generation of X-ray IFU instruments.

While in this work we focused on galaxies from the \textit{Magneticum} simulation, the obtained column densities for the major ions (\ovii and \oviii lines) for the massive and Milky Way-type galaxies are comparable with those obtained for the IllustrisTNG and EAGLE simulations \citep{nelson18,wijers20}. In addition, these simulations point out that the column densities only weakly depend on the radius, making the proposed approach ideal to study the CGM on the largest scales. The simulations also reveal that the column density of galaxies at the same stellar mass exhibits a significant galaxy-to-galaxy variation. This large variation is in agreement with emission studies, which also hint that galaxies exhibit significant variations at smaller radii (see Section \ref{sec:intro}). 
 
Finally, we mention that using the cumulative spectrum of CXB sources is not only ideal to study the large-scale CGM around galaxies, but other hot and low-density gas can also be traced. A prime example is the study of the warm-hot intergalactic medium, which has been traditionally explored with absorption studies using the observations of individual bright quasars. Since typical absorption studies of the WHIM are hampered by filament-to-filament variations, probing single sightlines with the population of CXB sources can yield a more comprehensive and less biased view of the baryon content and the characteristics of the baryons in the filamentary structure of the universe.  A follow-up study will explore the prospects of studying the WHIM in absorption with an X-ray IFU.

\section{Summary}
\label{sec:summary}

In this work, we explore the possibility to study the CGM on the largest scales in absorption using the cumulative spectrum of CXB sources. To this end, we utilize the hydrodynamical structure formation simulation, {\it Magneticum}, and build realistic mock images and construct column density maps. Our study and results can be summarized as follows. 

\begin{itemize}
    \item We constructed realistic mock images of four massive and five Milky Way-like galaxies that include the main X-ray emitting components, such as the large-scale CGM emission, the Milky Way foreground, and the population of CXB sources. We also generated column density maps for the major ions out to large radii. 
    \item We generated large sets of simulated spectra that include all the above-described components and probed whether absorption lines can be detected in the $(0.3-1)R_{\rm 200}$ and $(0.3-1.75)R_{\rm 200}$ regions for massive and Milky Way-type galaxies, respectively.  
    \item We established that the \ovii absorption line could be detected at the $3\sigma-6\sigma$ confidence level and the \oviii can be detected at the $1-3\sigma$ level around massive galaxies. The detection significance correlates with the median column densities and does not depend on small variations in the number or total luminosity of CXB sources within the extraction region. 
    \item Absorption lines around individual Milky Way-type galaxies remain either weakly detected or undetected. However, when co-adding the data from multiple galaxies, statistically significant detections can be obtained for the \ovii ($\sim6\sigma$) and \oviii ($\sim3\sigma$) ions. 
    \item We discuss that the observed absorption lines will not only allow probing of the baryon budget of galaxies but will also characterize the CGM at the largest scales. We also discuss that the ideal instrument to carry out the demonstrated analysis is a large-area and high-resolution microcalorimeter instrument.  
\end{itemize}

\smallskip

\begin{small}
\noindent
\textit{Acknowledgements.}
We thank the referee for the useful comments. \'A.B., R.P.K, C.J., and W.R.F acknowledge support from the Smithsonian Institution and the Chandra High Resolution Camera Project through NASA contract NAS8-03060. O.E.K is supported by the GA\v{C}R EXPRO grant No. 21-13491X. I.K. and K.D. acknowledge support by the COMPLEX project from the European Research Council (ERC) under the European Union’s Horizon 2020 research and innovation program grant agreement ERC-2019-AdG 882679. K.D. acknowledges support by the Deutsche Forschungsgemeinschaft (DFG, German Research Foundation) under Germany’s Excellence Strategy - EXC-2094 - 390783311. S.V.Z. acknowledges support by the DFG project nr. 415510302. The material is based upon work supported by NASA under award number 80GSFC21M0002.
\end{small}

\bibliographystyle{apj}
\bibliography{paper2.bib} 

\end{document}